\documentclass[10pt,journal,compsoc]{IEEEtran}
\ifCLASSOPTIONcompsoc
  \usepackage[nocompress]{cite}
\else
  \usepackage{cite}
\fi
\ifCLASSINFOpdf
\else
\fi

\usepackage[T1]{fontenc}
\usepackage[utf8]{inputenc}
\usepackage{graphicx}
\usepackage{subfigure}
\usepackage[hidelinks]{hyperref}
\usepackage{xspace}
\usepackage{xcolor}
\usepackage{amsmath}
\usepackage{amsfonts}
\usepackage{stmaryrd}
\usepackage{enumitem}
\usepackage{multirow}
\usepackage{booktabs}
\usepackage{amsthm}

\newcommand{\hide}[1]{}
\newcommand{\service}{Top Stories\xspace}
\newcommand{\model}{DiffuseGNN\xspace}
\newcommand{\modelat}{$\text{DiffuseGNN}_{\text{AA}}$\xspace}
\newcommand{\modeldt}{$\text{DiffuseGNN}_{\text{DA}}$\xspace}
\newcommand{\vpara}[1]{\vspace{0.07in}\noindent\textbf{#1 }}
\newcommand{\beq}[1]{\vspace{-0.1in}\begin{equation}#1\end{equation}\vspace{-0.1in}}
\usepackage[group-separator={,}]{siunitx}
\theoremstyle{definition}
\newtheorem{definition}{Definition}[section]

\begin{document}
%
\title{Understanding WeChat User Preferences and ``Wow'' Diffusion}
%
%
%
%

\author{Fanjin Zhang, Jie Tang$^{\star}$, \IEEEmembership{Fellow, IEEE}, Xueyi Liu, Zhenyu Hou, Yuxiao Dong, Jing Zhang, Xiao Liu, Ruobing Xie, Kai Zhuang, Xu Zhang, Leyu Lin, and Philip S. Yu, \IEEEmembership{Fellow, IEEE}
\IEEEcompsocitemizethanks{\IEEEcompsocthanksitem Fanjin Zhang is with the Department of Computer
Science and Technology, Tsinghua University and the WeChat Search Application Department, Tencent.
Xueyi Liu, Zhenyu Hou and Xiao Liu are with the Department of Computer
	Science and Technology, Tsinghua University, Beijing, 100084, China. 
	E-mail: \{zfj17, xueyi-li18, hzy17, liuxiao17\}@mails.tsinghua.edu.cn
	\IEEEcompsocthanksitem Jie Tang is with the Department of Computer Science and Technology, Tsinghua University, and Tsinghua National Laboratory for Information Science and Technology (TNList), Beijing, 100084, China. 
	E-mail: jietang@tsinghua.edu.cn, corresponding author
	\IEEEcompsocthanksitem Yuxiao Dong is with Microsoft Research, Redmond, WA 98052, USA 
	Email: ericdongyx@gmail.com
	\IEEEcompsocthanksitem Jing Zhang is with the Information School, Renmin University of China, Beijing, 100872, China. 
	Email: zhang-jing@ruc.edu.cn
	\IEEEcompsocthanksitem Ruobing Xie, Kai Zhuang, Xu Zhang and Leyu Lin are with the 
	WeChat Search Application Department, Tencent, Beijing, 100080, China. 
	Email: \{ruobingxie, terryzhuang, xuonezhang, goshawklin\}@tencent.com
	\IEEEcompsocthanksitem Philip S. Yu is with the Department of Computer Science, University of Illinois at Chicago, Chicago, IL 60607, USA.
	Email: psyu@uic.edu
}
\thanks{}
}

\markboth{IEEE TRANSACTIONS ON KNOWLEDGE AND DATA ENGINEERING, VOL. XX, NO. XX, February 2021}
{Shell \MakeLowercase{\textit{et al.}}: Bare Demo of IEEEtran.cls for Computer Society Journals}
%



\IEEEtitleabstractindextext{%
\begin{abstract}
WeChat is the largest social instant messaging platform in China,
with 1.1 billion monthly active users. 
``Top Stories'' is a novel friend-enhanced recommendation
engine
in WeChat, in which users
can read articles based on preferences of both their own and their friends. 
Specifically, when a user reads an article by opening it, the ``click'' behavior is private. 
Moreover, if the user clicks the ``wow'' button, (only) her/his direct connections will be aware of this action/preference. 
Based on the unique WeChat data, 
we aim to understand user preferences and ``wow'' diffusion
in \service
at different levels. 
We have made some interesting discoveries. 
For instance, the ``wow'' probability of one user is negatively correlated with the number of connected components that are formed by her/his active friends, but the click probability is the opposite. 
We further study to what extent users’ ``wow'' and click behavior can be predicted from their social connections. 
To address this problem, we present a hierarchical graph representation learning based model \model, which is capable of capturing the structure-based social observations discovered above.  
Our experiments show that the proposed method can significantly improve the prediction
performance 
compared with alternative methods.
\end{abstract}

}

\maketitle

\IEEEdisplaynontitleabstractindextext

%
\IEEEpeerreviewmaketitle

\IEEEraisesectionheading{\section{Introduction}\label{sec:introduction}}

\IEEEPARstart{I}{nformation} diffusion~\cite{rogers2010diffusion} has increasingly changed from offline to online these years.
There emerge many popular social applications, such as ``News Feed'' in Facebook and ``Top Stories'' in WeChat, which facilitate information diffusion greatly.
Central to information diffusion are the user trait and the social influence between users, which have attracted many researchers working on it~\cite{yang2015rain,granovetter1978threshold,goldenberg2001talk}.

Despite popular applications and extensive studies of information diffusion algorithms, it is still unclear about the inherent factors that result in different types of user feedback. 
First, how can user attributes, the correlations between users, and the local network structure influence user behavior? 
Second, what are the differences between various kinds of user feedback (such as ``click'', ``like'', and ``share'') w.r.t. the above factors?
Such problems are still largely unexplored and far from understood.

\begin{figure}[t]
	\centering
	\hspace*{0cm}
	\hspace{-0.05in}
	\includegraphics[width=7.5cm]{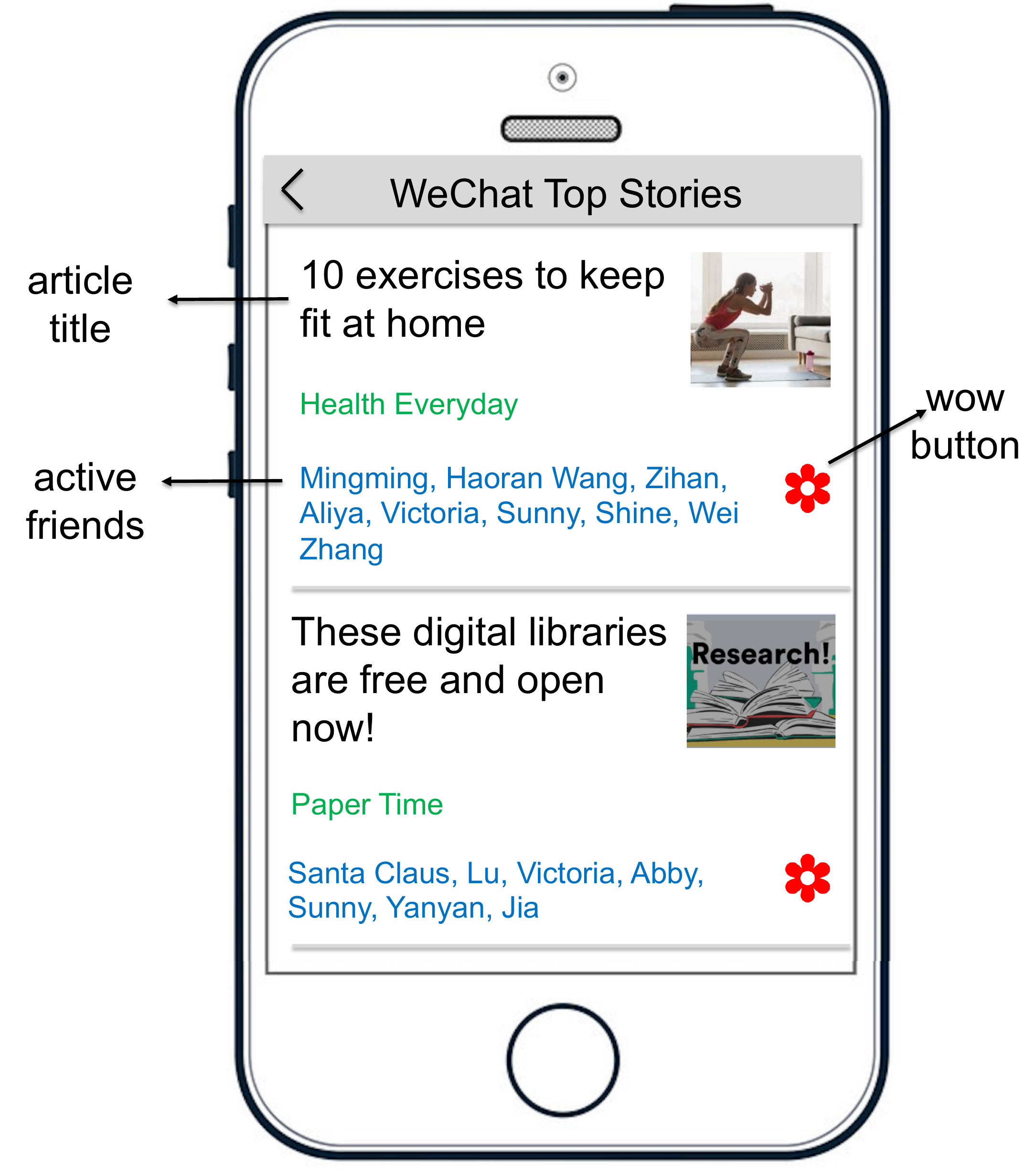}	
	\vspace{-0.1in}
	\caption{``Top Stories'' in WeChat. \textmd{Each user can view what friends ``wow''ed. If she/he also ``wow''s one article, it will be displayed to (only) her/his friends, which forms a diffusion process.}
	Here ``active friends'' means friends who ``wow''ed the corresponding articles.
	}
		\vspace{-0.3cm}
	\label{fig:wechat_wow_app}
\end{figure}

In this work, we 
study the  ``Top Stories'' service in WeChat --- the largest social instant messaging platform in China --- to 
understand user behavior in the specific social context.
In \service,
a user can see the articles ``wow''ed by her friends, which can be regarded as \textbf{share} plus \textbf{like}, and she can perform a ``wow'' or click action on each article as well. 
An illustrative example of WeChat's \service service is shown in Figure \ref{fig:wechat_wow_app}, in which each user is shown articles that her friends ``wow''ed as well as those friends' names, and she can click on the articles or also ``wow'' them.   
Herein, we aim at understanding users' ``wow'' and click behavior from different aspects, including user demographics, dyadic and triadic correlations, and users' ego network structures. 

The problem in this paper is related to social influence locality~\cite{zhang2013social}, which targets quantifying how user behavior is influenced by other users in ego networks,
and is more broadly related to social influence\cite{tang2009social} and information diffusion~\cite{matsubara2012rise,Saito2008Influence}.
Most methods address the social influence locality problem by using hand-crafted user features and network features to predict user behaviors~\cite{zhang2013social,lin2017detecting}. Recently, Qiu et al.~\cite{qiu2018deepinf} propose to use graph attention networks (GAT) to learn user proximity in ego networks. 
Wang et al.~\cite{wang2020social} further consider topological features via the Weisfeiler-Lehman (WL) algorithm~\cite{douglas2011weisfeiler} to predict user behavior for in-feed advertising. 
However, they haven't dug into studying potential 
factors in different granularities, such as
user demographics and ego network properties,
based on which better prediction models can be designed.

Through the study, we first reveal several intriguing discoveries that impact user behavior at different levels. 
Based on these discoveries, we develop a unified framework to predict users' ``wow''  and click behavior by modeling user attributes, dyadic correlations, and ego network structures together.

To highlight several of our key findings, for user demographics, we find that 
users' ``wow'' and click behavior vary by gender and age, and the patterns become complicated when cross-attribute factors are considered.
For dyadic correlations, 
users are likely to behave differently when their active friends are structural holes and opinion leaders.
Considering ego network properties, 
both ``wow'' probability and click probability are strongly correlated with the number of connected components formed by users' active friends,
but they have the opposite patterns.
This correlation becomes stronger when the ego network is cleaned.

Based on these interesting discoveries, we further study to what extent users' behavior can be predicted from their social connections and attributes. 
To this end, 
we propose a hierarchical graph representation learning based model called \model. 
Our model is closely related to and motivated by the insights from data, which is different from many neural network based methods.
Specifically, first, 
\textbf{to model cross-attribute factors for users' different attributes as the analysis in Section \ref{subsec:user_demo}} (such as user embeddings and demographics), we adopt the factorization machine technique to generate second-order  features to model feature interactions for each individual.
Second, \textbf{to remove noise in the ego networks as shown in the Section \ref{sec:ego-net-property} analysis}, \model propagates initial user features in the modulated spectral domain,
to generate user embeddings based on cleaned ego networks.
Third, \textbf{to model dyadic correlations as the analysis in Section \ref{subsec:user-relation} shows},
we adopt a new graph attention mechanism to model feature interactions between neighbors.
Fourth, \textbf{to model the connected components --- the hierarchical structure of the ego networks as in the Section \ref{sec:ego-net-property} analysis},
we generate hierarchical representations of ego networks by clustering nodes together and learning on the coarsened graphs iteratively.
We evaluate the proposed method on a WeChat \service dataset and a public Weibo dataset. 
Our experiments show that the proposed solution 
can consistently outperform 
alternative methods. 
\section{Background --- Top Stories Dataset}
Different from other news feed systems, 
Top Stories in WeChat recommend to a user articles favored (``wow''ed) by her/his friends.
As shown in Figure~\ref{fig:wechat_wow_app}, the articles recommended to the current user are two articles favored by her active friends. The user can choose to click the ``wow'' button so that her friends will also be informed of her choice. 
In this way, the ``wow''
essentially plays an implicit diffusion role.
On the other hand, the user can also click to view the full content of the article, or simply ignore it. 
The dataset used in this paper is a subset of data sampled randomly from \service.
The approval from users were developed prior to the start of data collection.
It consists of three parts: 
1) a social network $G = \{U, E\}$, where $U$ is a set of users, and $E$ represents a set of edges recording friendships between users;
2) user attributes $C$ including users' gender, age, regions, and so on; 
and 3) the interaction between users and articles $L=\{(u, d, ts, is\_like, is\_click, af(u, d, ts)) | u \in U, d \in \mathcal{D}\}$, where $u$ is the ego user, $d$ is a displayed article in article set $\mathcal{D}$, $ts$ is the 
timestamp, $is\_like$ and $is\_click$ are whether $u$ ``wow''s and clicks $d$, 
$af(u, d, ts)$ is $u$'s active friends who ``wow''ed $d$ before timestamp $ts$.
Note that an interaction can be  represented by a triplet $(u, d, ts)$.
To avoid over-fitting, we further select a subset of the data by first extracting users who performed at least ten interactions (``wow'' or click), and then extracting these users' friendship networks and their attributes. 
The final dataset contains \num{48084772} users, \num{61252317} articles, and \num{7459660092} interactions.
All data are preprocessed via data masking to protect user privacy.

To start with our analysis, we first introduce several definitions which will be used later.

\theoremstyle{definition}
\begin{definition}{\bf Connected Components (CC)\footnote{\url{https://en.wikipedia.org/wiki/Component_(graph_theory)}}.}
In graph theory, a connected component of an undirected graph $G$ is an induced subgraph in which any two vertices are connected to each other by paths, and which is connected to no additional vertices in the rest of the graph. We let \textbf{CC} denote the acronym of \textit{connected components} and \textbf{\#CC} denote the number of connected components in a graph.
\end{definition}

\theoremstyle{definition}
\begin{definition}{\bf Structural Hole (SH)~\cite{burt2004structural}.}
A “structural hole” is a term for recognizing a missing bridge in a graph.
Wherever two or more groups fail to connect, one can argue that there is a structural hole, a gap waiting to be filled.
We let \textbf{SH} denote the acronym of \textit{structural hole}.
\end{definition}

\theoremstyle{definition}
\begin{definition}{\bf Ego network and $\tau$-ego network.}
Ego network $G_u=\{U_u, E_u\}$ is a subgraph of a static social network $G$ centered at the focal node (``\textit{ego}''), where $U_u$ is the node set consisting of the ego and its first-order neighbors, 
$E_u$ is the edge set containing edges between nodes in $U_u$ in the original graph $G$.
\textbf{$\boldsymbol{\tau}$-ego network} $G_u^{\tau} = \{V_u^{\tau}, E_u^{\tau}\}$ is a subgraph induced by $u$ and $u$'s $\tau$-degree friends, $V_u^{\tau}$ is the node set of the subgraph $G_u^{\tau}$ 
and $E_u^{\tau}$ is the edge set of $G_u^{\tau}$.
\end{definition}

\theoremstyle{definition}
\begin{definition}{\bf Active Friends.}
	In this article, we define active friends as the friends who performed ``wow'' action on an article.
	In WeChat Top Stories, if a user ``wow''ed an article, his/her friends will be informed about it.
	However, users' click behavior will not be shown directly to friends. As shown in Figure \ref{fig:wechat_wow_app}, the names shown below the articles are the friends who performed a ``wow'' action.
\end{definition}

\theoremstyle{definition}
\begin{definition}{\bf Active Rate.}
We define \textbf{``active rate''} to refer to both ``wow'' probability and click probability.
\end{definition}
\section{Analysis and Discoveries}
Based on abundant user behavior data in \service,
we investigate how users' ``wow'' and click behavior correlates with three aspects:
(1) user demographics, (2) dyadic and triadic correlations, and (3) ego network properties.
Next, we present these analysis results one by one.

\subsection{User Demographics}
\label{subsec:user_demo}

\begin{table}
	\caption{User activity w.r.t. gender.}
	\label{tb:user-gender-act}
	\centering
	\begin{tabular}{c|c|c}
		\hline
		\textbf{Gender} & \textbf{``Wow'' prob.} & \textbf{Click prob.} \\
		\hline
		Male  & 1.17\% & \textbf{10.62}\%\\
		\hline
		Female & \textbf{1.19\%} & 9.86\%
		\\
		\hline
	\end{tabular}
\end{table}

\begin{figure}[t]
	\centering
	\includegraphics[width=5.5cm]{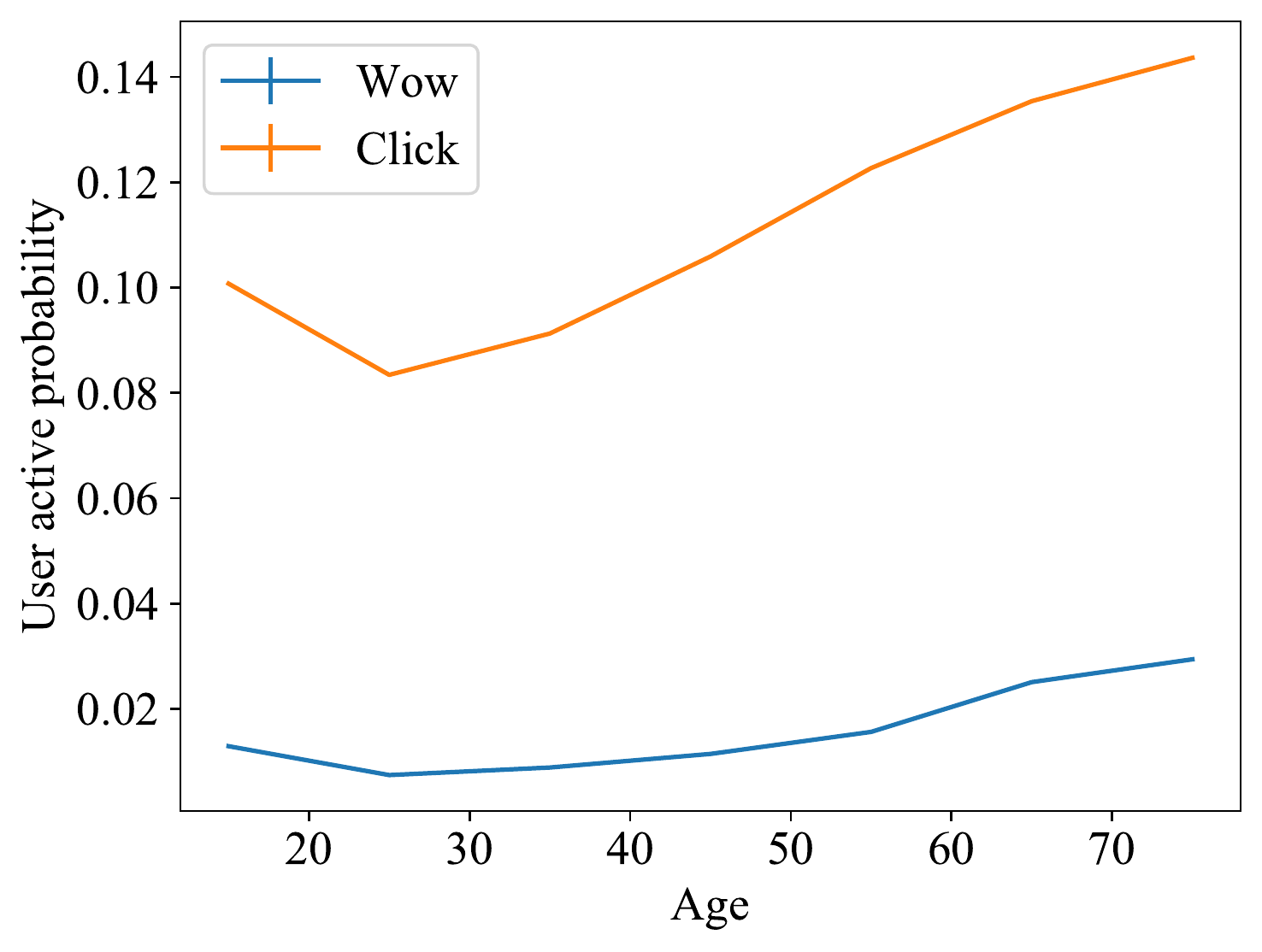}	
	\vspace{-0.5cm}
	\caption{``Wow'' and click probability w.r.t. user age.}
	\label{fig:user-age-act}
	\vspace{0.5cm}
\end{figure}

\begin{figure}[t]
	\centering
	\hspace*{0cm}
	\hspace{-0.05in}
	\mbox{
		\subfigure[``Wow'' Behavior]{
			\includegraphics[width=4cm]{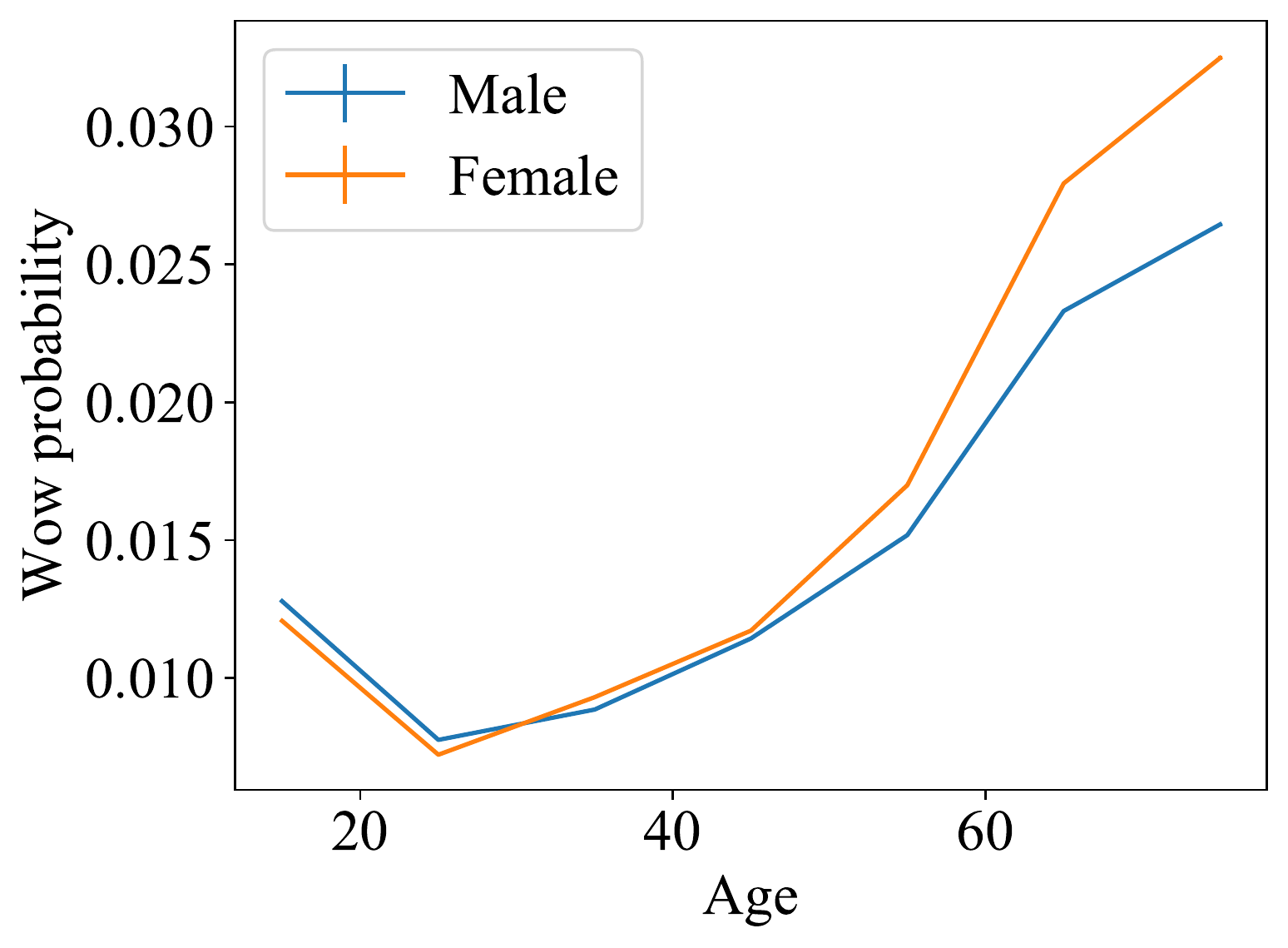}	
			\label{figsub:self-gender-age-wow}
		}
		\hspace{0.1in}
		\subfigure[Click Behavior]{
			\includegraphics[width=4cm]{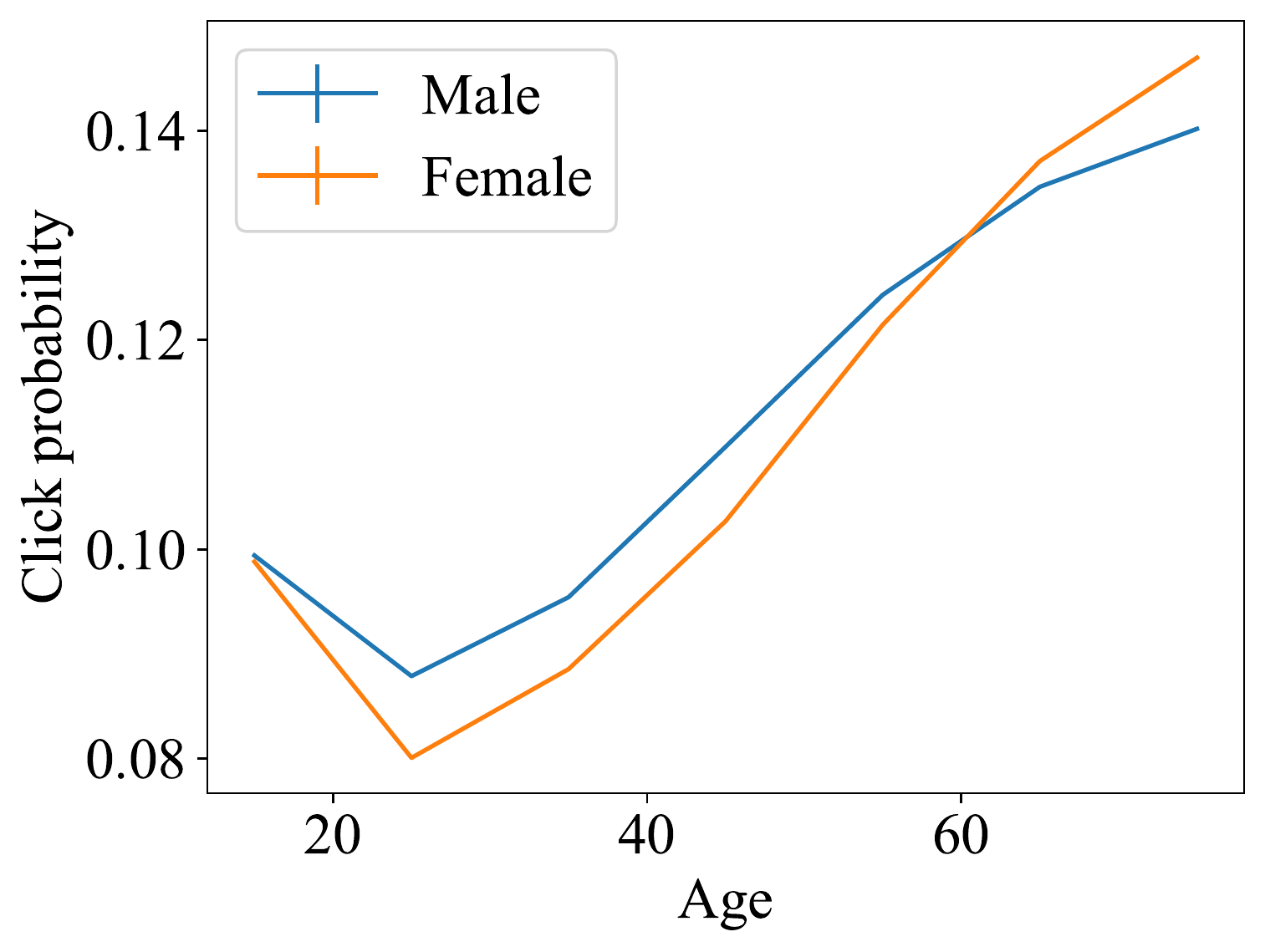}	
			\label{figsub:self-gender-age-click}
		}
	}
	\vspace{-0.13in}
	\caption{
		User ``wow'' and click probability w.r.t. users' gender and age.
	}
	\label{fig:self-gender-age-act}
\end{figure}

Table \ref{tb:user-gender-act}, Figure \ref{fig:user-age-act}, and Figure \ref{fig:self-gender-age-act} show the probability that people of different gender and age ``wow'' or click articles in \service. 
From Table \ref{tb:user-gender-act}, we observe that males' click probability is clearly higher than females',
while females' ``wow'' probability is a little bit higher than males'. The reason might be that males tend to consume content, but females are more active in social circles. 
Regarding age, 
the patterns are very interesting. According to our intuition, young generations (users in their 20s and 30s) are the most active users in our online social circles. However, in Figure \ref{fig:user-age-act}, the ``wow'' and click probability of the 20s and 30s is the lowest among all ages. 
We infer that young people might be too busy to look at the articles in detail, or
their ``reverse psychology'' reacts to the recommended articles.

Moreover, when we consider both gender and age attributes, the patterns become different again. In Figure \ref{fig:self-gender-age-act}, 
we find that for people younger than 20, males are more active than females. 
However, there is a reversion for both ``wow'' and click behavior, but at different split points (the 40s for ``wow'' and the 60s for click), indicating that older female users are more active than older male users.
The result demonstrates that the cross-attribute factor is more complicated.

\subsection{Dyadic and Triadic Correlations}
\label{subsec:user-relation}
In this subsection, we consider dyadic and triadic correlation factors. To eliminate other influence factors, we consider interactions with only one friend who ``wow''ed the article for dyadic correlations, and exactly two active friends for triadic correlations. 
We analyze these correlations from two views: demographic attributes and social roles.

\begin{table}
	\caption{Dyadic correlations w.r.t. user gender and friend gender.}
	\label{tb:user-pairwise-gender-act}
	\centering
	\begin{tabular}{c|c|c|c}
		\hline
		\textbf{User} & \textbf{Friend} & \textbf{``Wow'' prob.} & \textbf{Click prob.} \\
		\hline
		Male  & Male & 0.97\% & \textbf{11.19}\%\\
		\hline
		Male & Female & \textbf{1.01\%} & 9.69\%\\
		\hline
		Female & Male & 0.93\% & 9.11\% \\
		\hline
		Female & Female & \textbf{1.06\%} & \textbf{10.33\%}
		\\
		\hline
	\end{tabular}
\end{table}

\begin{figure}[t]
	\centering
	\includegraphics[width=7cm]{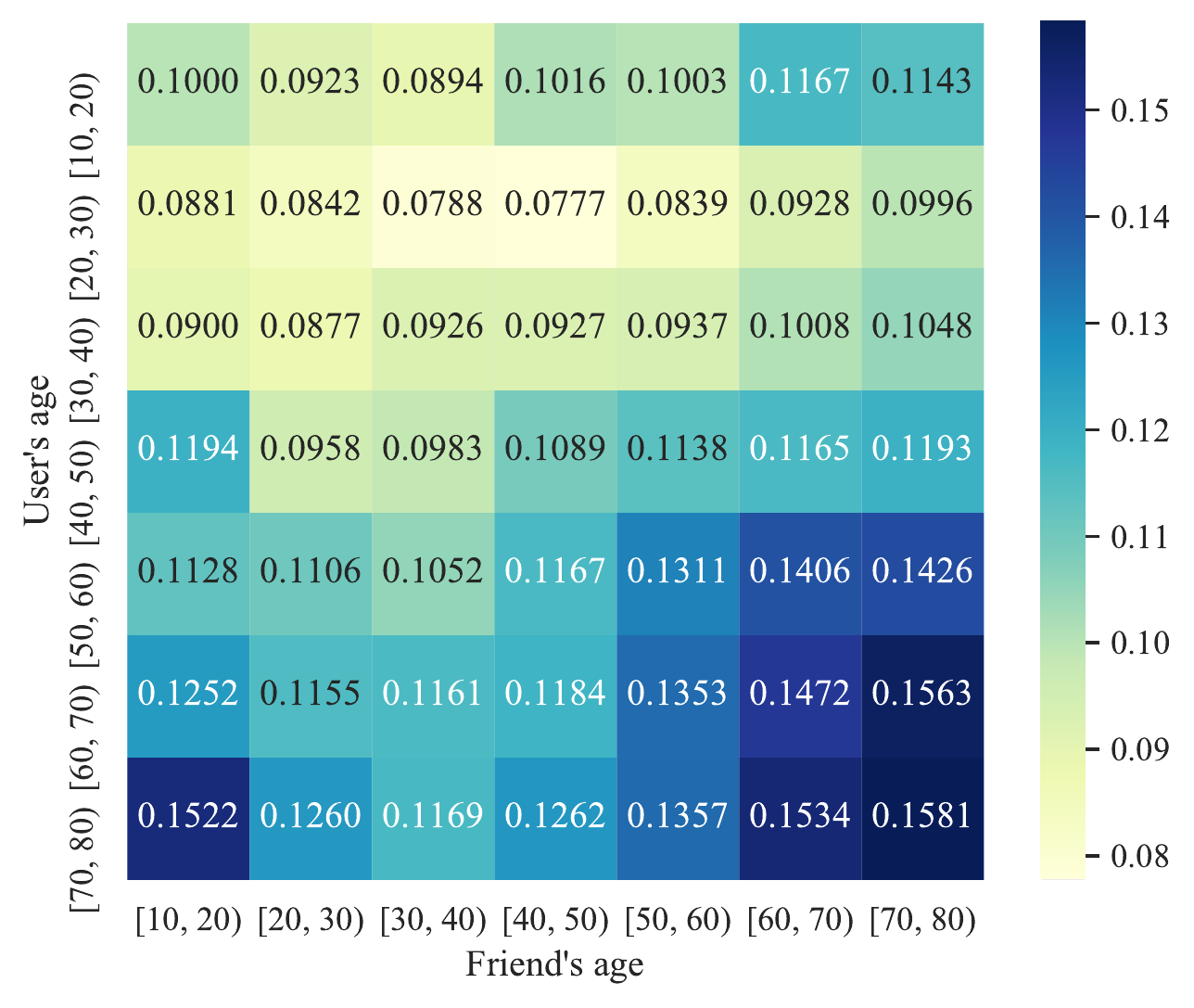}
	\vspace{-0.5cm}
	\caption{Users' ``wow'' probability w.r.t. user age and friend age.}
	\label{fig:user-pairwise-age-act}
	\vspace{0.5cm}
\end{figure}

\begin{table}
	\caption{Dyadic correlations w.r.t. the distance between the user and the friend.}
	\label{tb:user-pairwise-poi-act}
	\centering
	\begin{tabular}{c|c|c}
		\hline
		\textbf{User}  & \textbf{``Wow'' prob.} & \textbf{Click prob.} \\
		\hline
		All & 1.01\% & 10.24\%\\
		\hline
		Same province & 1.05\% & 10.65\% \\
		\hline
		Same city & 1.08\% & 10.85\% \\
		\hline
		Same district & \textbf{1.19\%} & \textbf{11.27\%}
		\\
		\hline
	\end{tabular}
\end{table}

\vpara{Dyadic Correlations w.r.t. Demographic Attributes.} Table \ref{tb:user-pairwise-gender-act} shows the users' active rate concerning dyadic correlations between ego users' gender and friends' gender. We observe that for click behavior, when friends' gender is the same as ego users', the ego users' click probability is higher, which can be explained by the homophily of their interests. However, for ``wow'' behavior, users are more likely to ``wow'' an article when their active friends are females.

In view of ages, Figure \ref{fig:user-pairwise-age-act} visualizes users' ``wow'' probability w.r.t. dyadic correlations between users' ages and friends' ages. We have several interesting discoveries. First, when users are young (< 40 years old), they are more influenced by their older friends than friends of the same age group. 
Second, older users are highly influenced by their friends of the same age group. Meanwhile, they also care about what young generations ``wow''ed, which shows cross-generation care, such as parents to children, managers to subordinates, etc. 
The pattern of click behavior is omitted here since it is similar to that of ``wow'' behavior.

Talking about the region, we also consider the distance between users and their friends (incorporating users' regions). Table \ref{tb:user-pairwise-poi-act} shows the users' active rate w.r.t. the distance between the user and the active friend.
We see that when the geographic distance between the ego user and the friend is closer, the ``wow'' probability and click probability of the ego user is higher, 
which shows the existence of interest homophily w.r.t. user region.

\begin{table}
	\caption{Dyadic correlations w.r.t. users' and friends' social roles. OU: ordinary user; OL: opinion leader.}
	\label{tb:user-pairwise-pr-act}
	\centering
	\begin{tabular}{c|c|c|c}
		\hline
		\textbf{User} & \textbf{Friend} & \textbf{``Wow'' prob.} & \textbf{Click prob.} \\
		\hline
		OU  & OU & \textbf{1.06\%} & \textbf{10.65}\%\\
		\hline
		OU & OL & 0.66\% & 8.11\%\\
		\hline
		OL & OU & \textbf{0.92\%} & \textbf{8.57\%} \\
		\hline
		OL & OL & 0.64\% & 7.49\%
		\\
		\hline
	\end{tabular}
\end{table}

\vpara{Dyadic Correlations w.r.t. Social Roles.}
We also study dyadic correlations between users' social roles and friends' social roles. Here social roles refer to users' roles in the social network or ``wow'' diffusion network. Table \ref{tb:user-pairwise-pr-act} shows the users' active rate w.r.t. dyadic correlations of different social roles: opinion leaders (OL) and ordinary users (OU). We find opinion leaders in social networks by running the PageRank algorithm~\cite{page1999pagerank} on the user diffusion network and then regard users with top 1\% PageRank scores as opinion leaders. Surprisingly, we find users' ``wow'' and click probability is higher when their active friends are not opinion leaders. 
For ``wow'' behavior, the reasons might be that if an opinion leader has ``wow''ed one article, the ego user is less willing to publicize it because many users connected to opinion leaders probably have already known it. For click behavior, we think perhaps users browse \service mainly for recreation, as WeChat is a social instant messaging platform for friends and acquaintances. Therefore, users might care more about what similar friends are interested in, rather than what opinion leaders pay attention to.

\begin{table}
	\caption{Dyadic correlations w.r.t. users' and friends' social roles. OU: ordinary user; SH: structural hole.}
	\label{tb:user-pairwise-sh-act}
	\centering
	\begin{tabular}{c|c|c|c}
		\hline
		\textbf{User} & \textbf{Friend} & \textbf{``Wow'' prob.} & \textbf{Click prob.} \\
		\hline
		OU  & OU & 0.99\% & 10.26\%\\
		\hline
		OU & SH & \textbf{1.38\%} & \textbf{12.43\%}\\
		\hline
		SH & OU & 2.34\% & \textbf{10.28\%} \\
		\hline
		SH & SH & \textbf{3.58\%} & 10.16\%
		\\
		\hline
	\end{tabular}
\end{table}

Table \ref{tb:user-pairwise-sh-act} also shows the users' active rate w.r.t. dyadic correlations of different social roles: structural holes (SH) and ordinary users (OU).
Here we find cut points in the users’ friendship
network using the Tarjan~\cite{tarjan1972depth} algorithm to approximate structural hole users.
Clearly, users' ``wow'' behavior is highly influenced when their friends are structural holes, which demonstrates that structural holes are critical in the information diffusion process. Also, the ``wow'' and click probability of ordinary users is higher if their active friends are structural holes.
For users who are structural holes, their click probability is higher when their friends are not structural holes, but the difference is not very significant.

\vpara{Triadic Correlations w.r.t. Demographic Attributes.}
Here we study how triadic correlations --- more complex factors --- would influence ego users' behavior. 
For triadic correlations, we consider interactions with exactly two friends who ``wow''ed the article. We analyze the demographics (gender, age, and region) of ego users and their friends.

\begin{figure}[t]
	\centering
	\includegraphics[width=8cm]{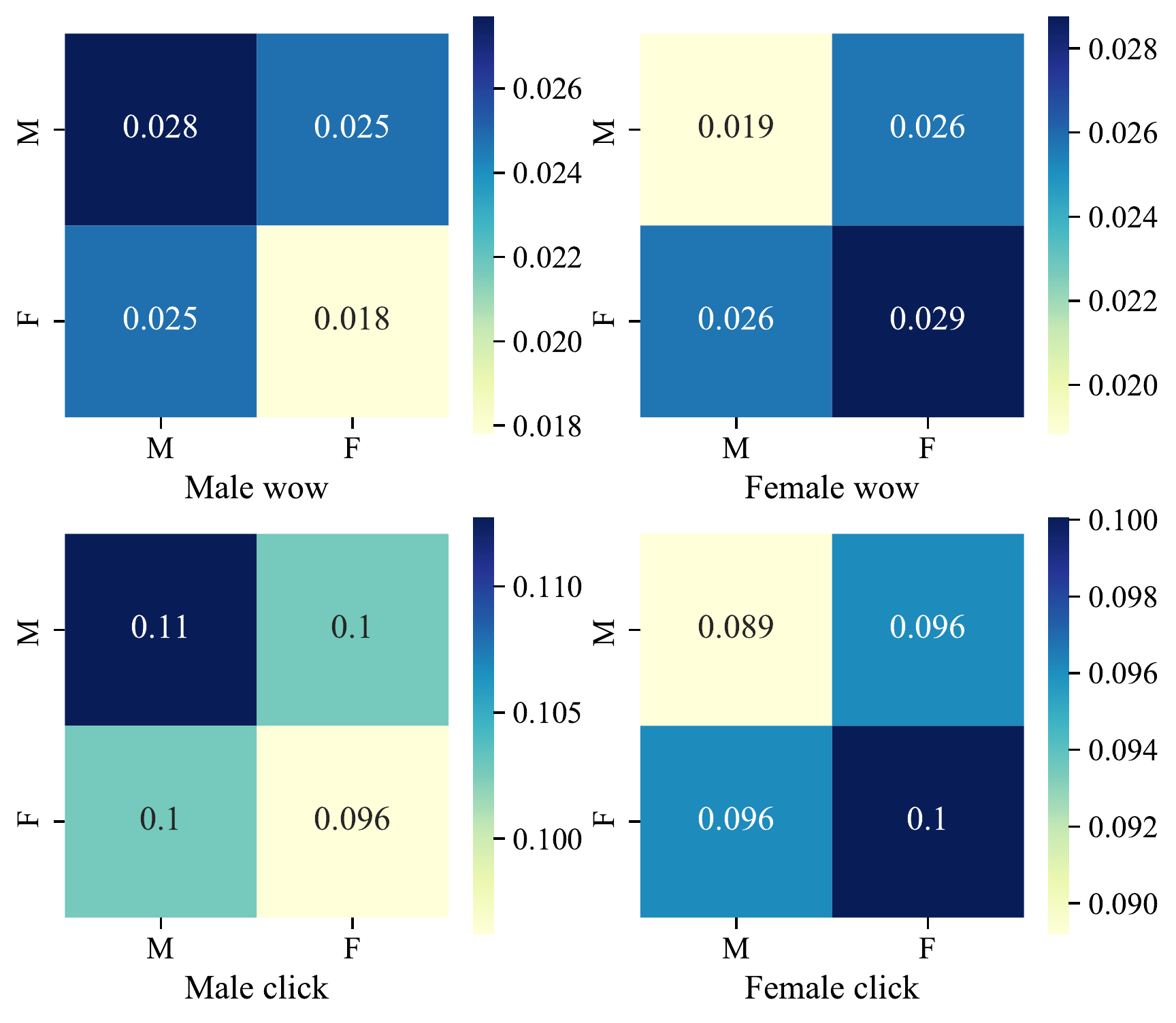}
	\vspace{-0.5cm}
	\caption{Users' ``wow'' and click probability w.r.t. user's gender and friends' gender. M: Male, F: Female.}
	\label{fig:user-triangle-gender-act}
	\vspace{0.5cm}
\end{figure}

\hide{

\begin{figure}[t]
	\centering
	\includegraphics[width=8.5cm]{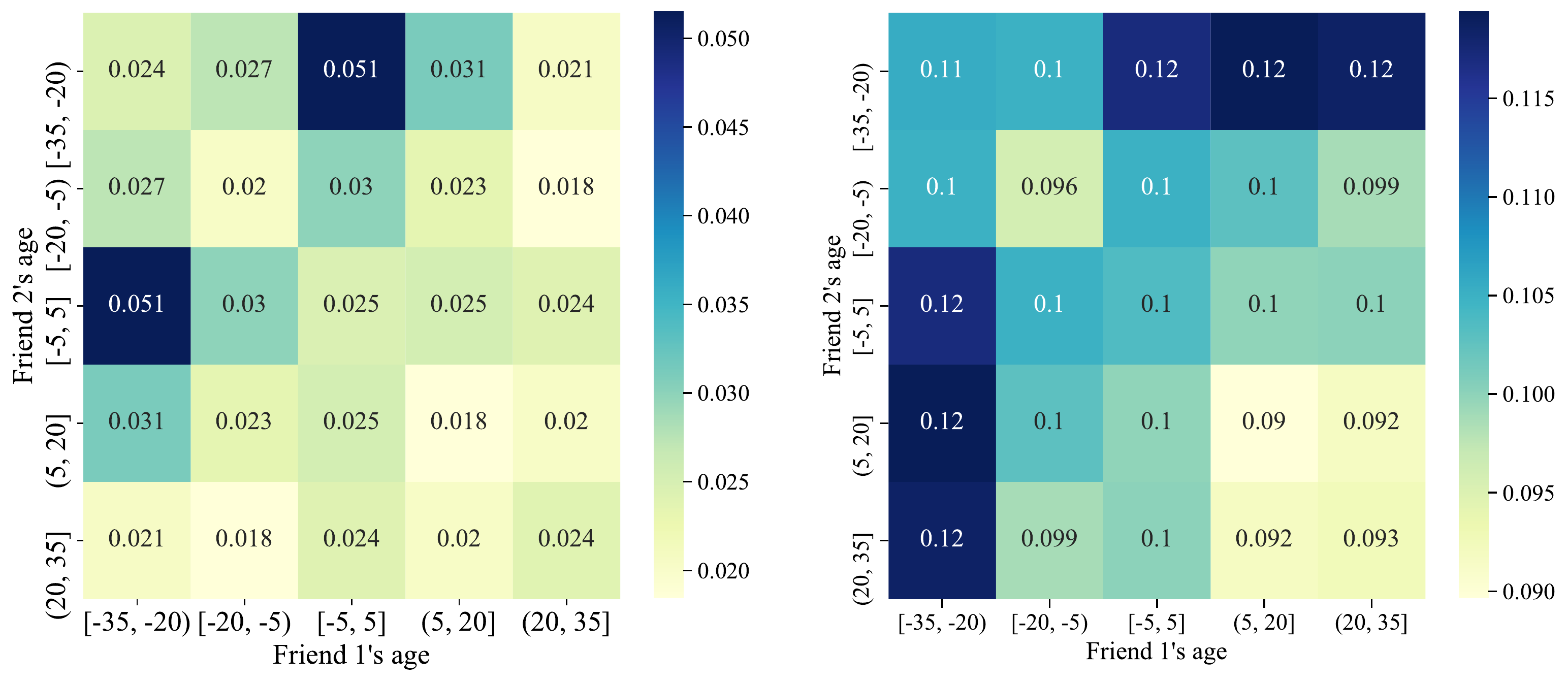}
	\vspace{-0.5cm}
	\caption{Users' ``wow'' and click probability w.r.t. the user's gender and friends' gender.}
	\label{fig:user-triangle-age-act}
	\vspace{0.5cm}
\end{figure}

}

\begin{figure}[t]
	\centering
	\hspace*{0cm}
	\hspace{-0.05in}
	\mbox{
		\subfigure[``Wow'' Behavior]{
			\includegraphics[width=4.1cm]{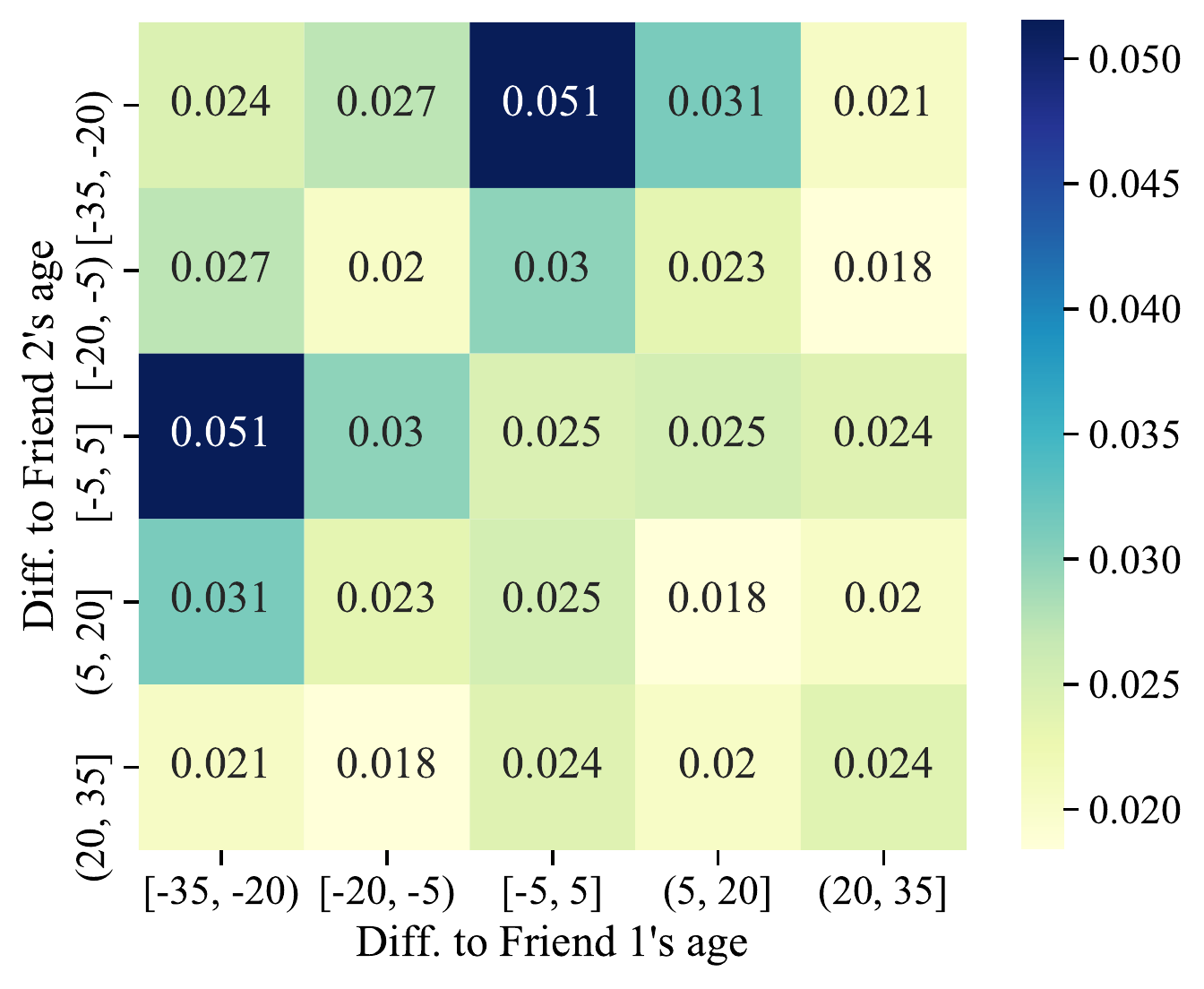}	
			\label{figsub:user-triangle-age-like}
		}
		\hspace{0.1in}
		\subfigure[Click Behavior]{
			\includegraphics[width=4.1cm]{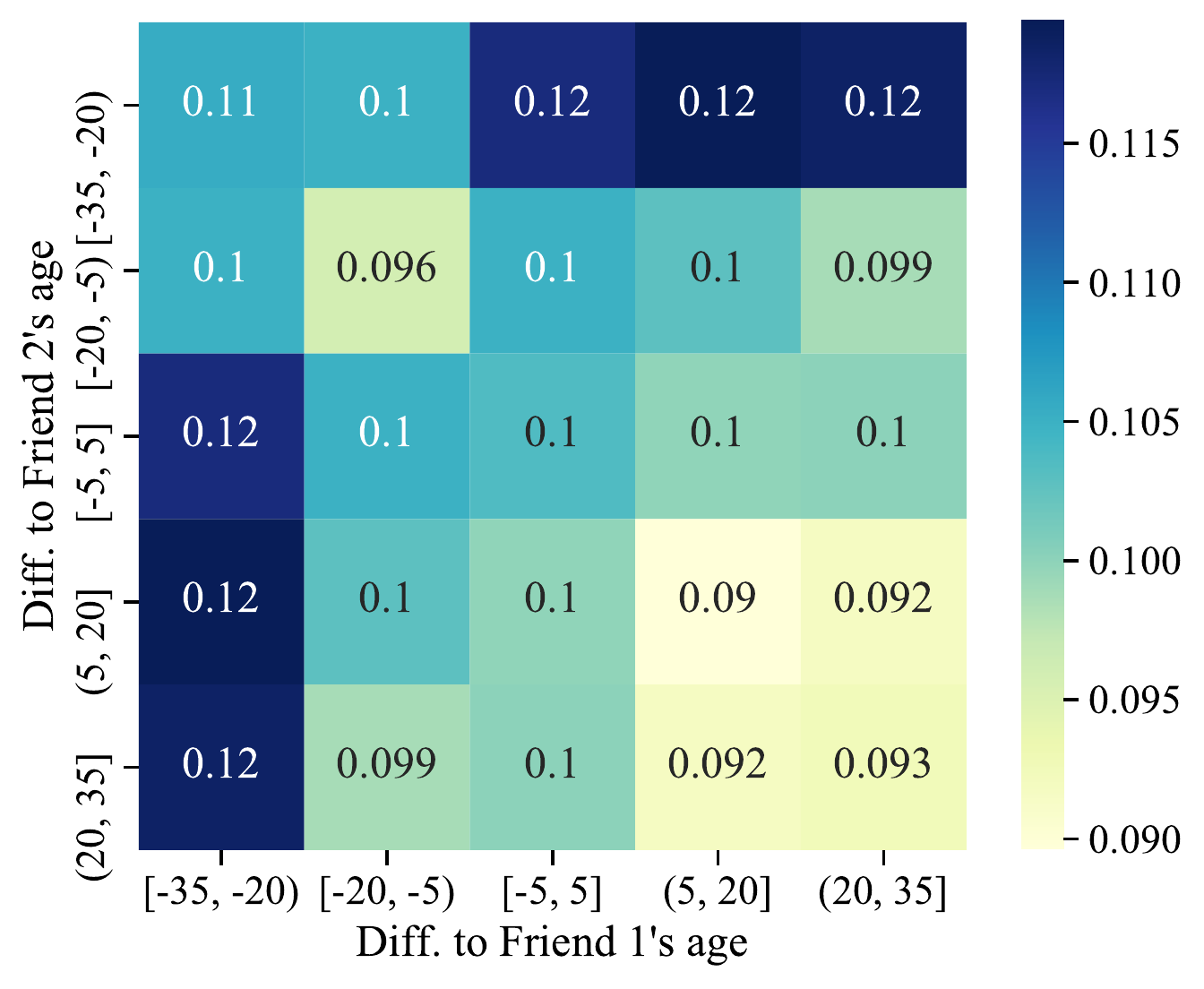}	
			\label{figsub:user-triangle-age-click}
		}
	}
	\vspace{-0.13in}
	\caption{
		User ``wow'' and click probability w.r.t. the differences between user age and friend age. Here the difference is calculated by ``the age of the friend minus the age of the ego user''.
	}
	\label{fig:user-triangle-age-act}
\end{figure}

\begin{figure}[t]
	\centering
	\hspace*{0cm}
	\hspace{-0.05in}
	\mbox{
		\subfigure[``Wow'' Behavior]{
			\includegraphics[width=4.1cm]{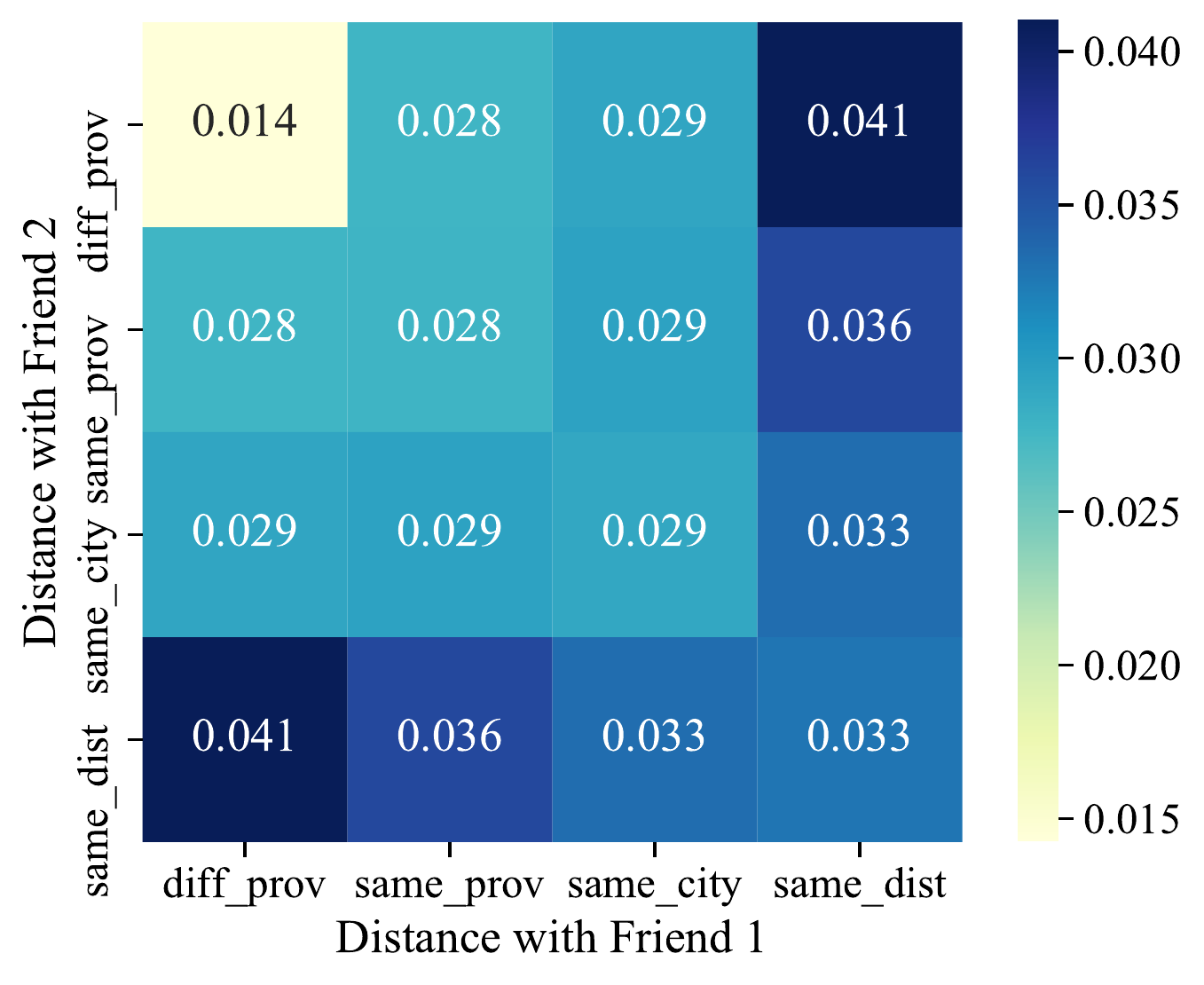}	
			\label{figsub:user-triangle-poi-like}
		}
		\hspace{0.1in}
		\subfigure[Click Behavior]{
			\includegraphics[width=4.1cm]{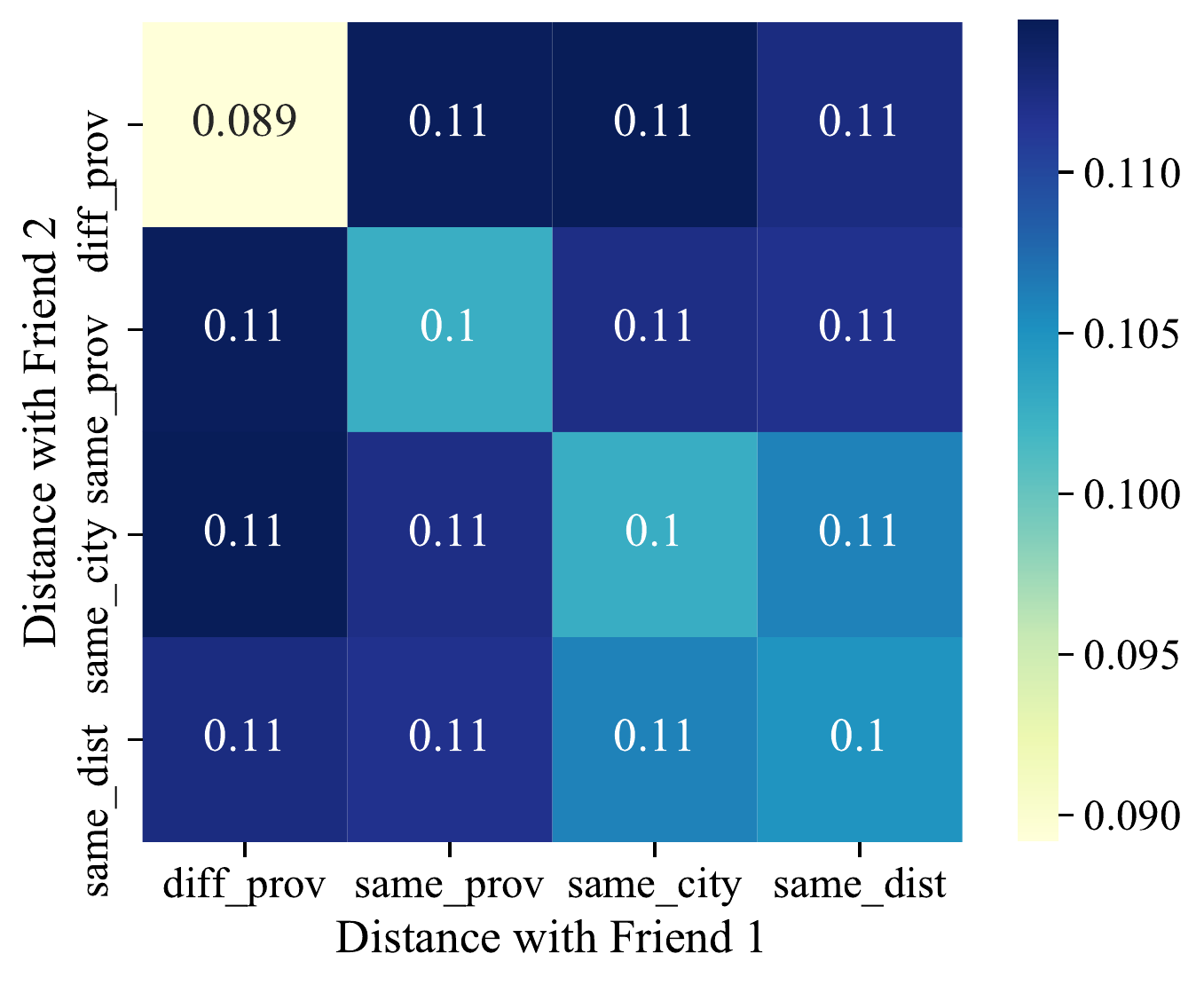}	
			\label{figsub:user-triangle-poi-click}
		}
	}
	\vspace{-0.13in}
	\caption{
		User ``wow'' and click probability w.r.t. the distance between the ego user and the friends.
		Here ``diff\_prov” means different
		provinces, ``same\_prov” means the same provinces, ``same\_city”
		means the same cities, and ``same\_dist” means the same districts.
	}
	\label{fig:user-triangle-poi-act}
\end{figure}

\hide{

\begin{figure}[t]
	\centering
	\includegraphics[width=8.5cm]{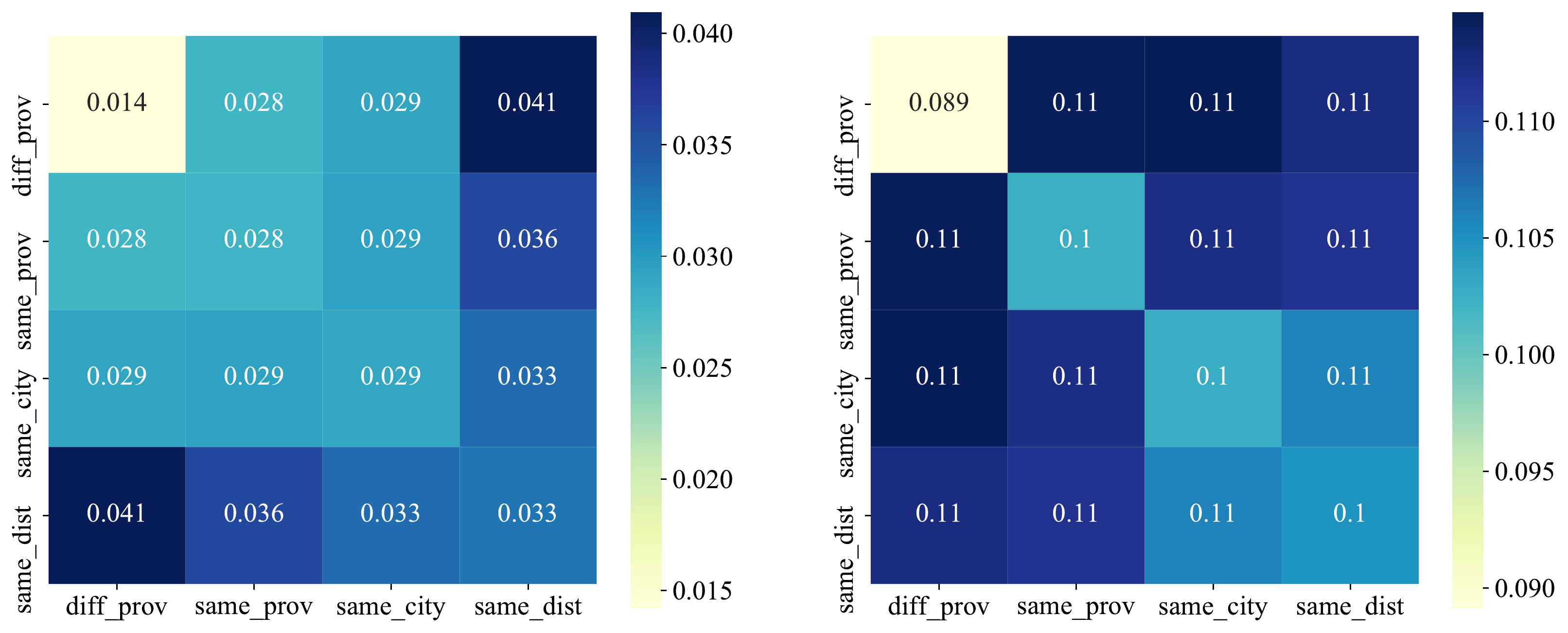}
	\caption{Users' ``wow'' and click probability w.r.t. the distance between the ego user and the two active friends.
	Here ``diff\_prov” means different
provinces, ``same\_prov” means same provinces, ``same\_prov”
means same cities and ``same\_dist” means same districts.}
	\label{fig:user-triangle-poi-act}
	\vspace{0.5cm}
\end{figure}

}

Figure \ref{fig:user-triangle-gender-act} shows the users' active
rate with respect to the triadic correlations between the ego user's gender and his/her two friends’ gender. From the figure, we observe consistent patterns of ``wow'' and click behaviors. If the two friends’ gender is
the same as the ego user’s gender, the ego users' active rate is the
highest. Again, this implies a high degree of gender homophily.

Furthermore, Figure \ref{fig:user-triangle-age-act} shows the users' active rate w.r.t. the difference between the ego user's age and the two friends' ages.
We discover that if one friend is of the same age group and 
the other friend is younger than the ego user, the active rate of the ego
user is high. 
Furthermore, the color in the left top is darker than
that of the right bottom, which demonstrates that old users pay more
attention to young users compared with young users attending to
old users.

Additionally, Figure \ref{fig:user-triangle-poi-act} shows the users' active rate w.r.t. the distance between the ego user and the two active friends.
Our intuition is that if the distance between the ego user and her two friends is
closer, the ego users' active rate can be higher. 
However, actually, if one
friend is nearby, and the other friend is more distant from the ego user, the
ego user will be more active. This kind of “attribute diversity” may
provide evidence that the ``wow''ed articles are acknowledged
by various users.

\subsection{Ego Network Properties}\label{sec:ego-net-property}
In this subsection, we study the correlation between users' activity and their ego network properties. 
To be precise,
the ego network is defined as the induced subgraph of users' active friends.
We find that users' online behaviors (click and ``wow'') are strongly influenced by their friend circles (users in their ego networks).
We study ego network properties from three aspects: the number of friends in the ego network, the number of connected components (\#CC) in the ego network, \#CC in the cleaned ego network (k-core subgraph).

\hide{

\begin{figure}[t]
	\centering
	\includegraphics[width=8cm]{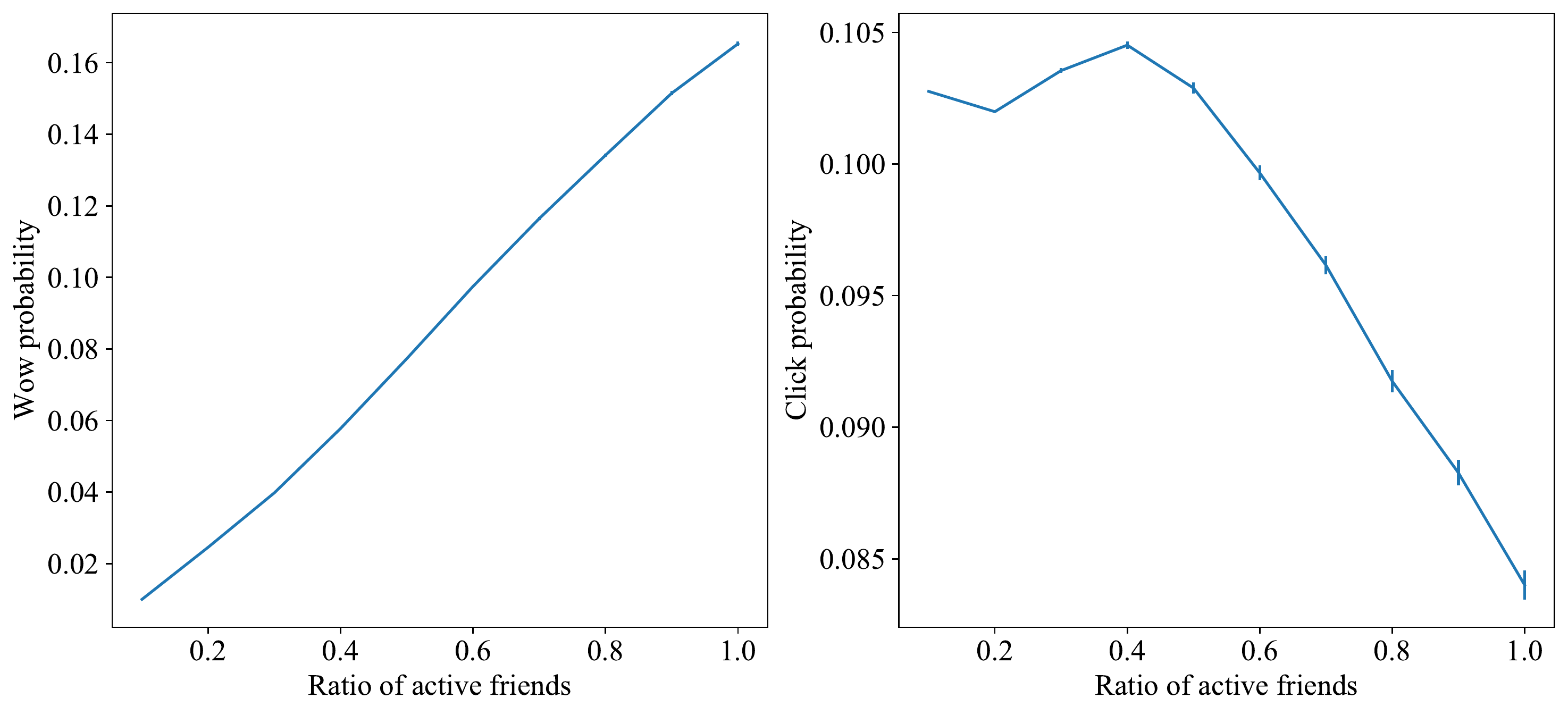}
	\vspace{-0.5cm}
	\caption{Influence v.s. \#active friends}
	\label{fig:influencefriend}
	\vspace{0.5cm}
\end{figure}

}

\begin{figure}[t]
	\centering
	\hspace*{0cm}
	\hspace{-0.05in}
	\mbox{
		\subfigure[``Wow'' Behavior]{
			\includegraphics[width=4.1cm]{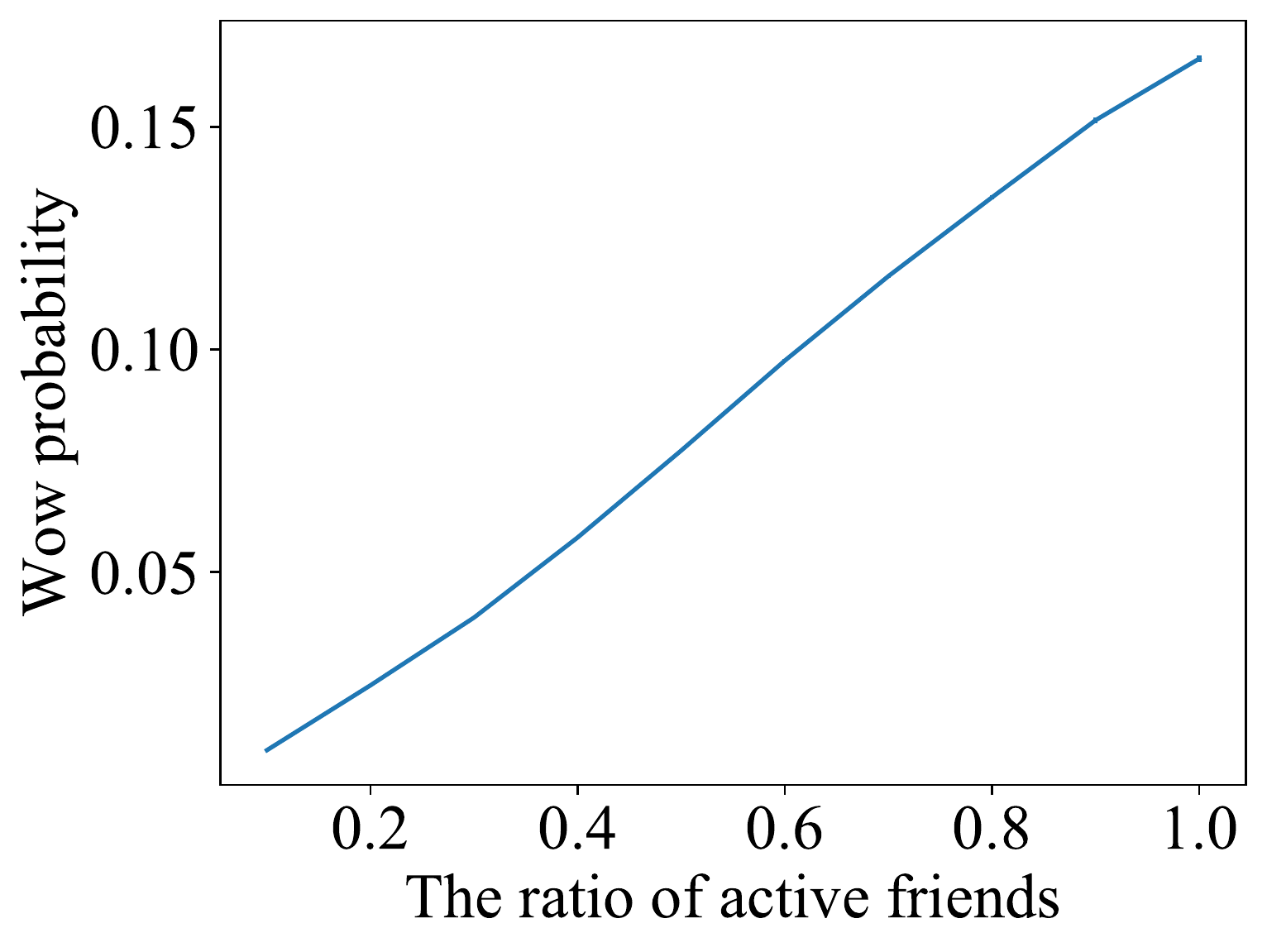}	
			\label{figsub:user-triangle-poi-like}
		}
		\subfigure[Click Behavior]{
			\includegraphics[width=4.1cm]{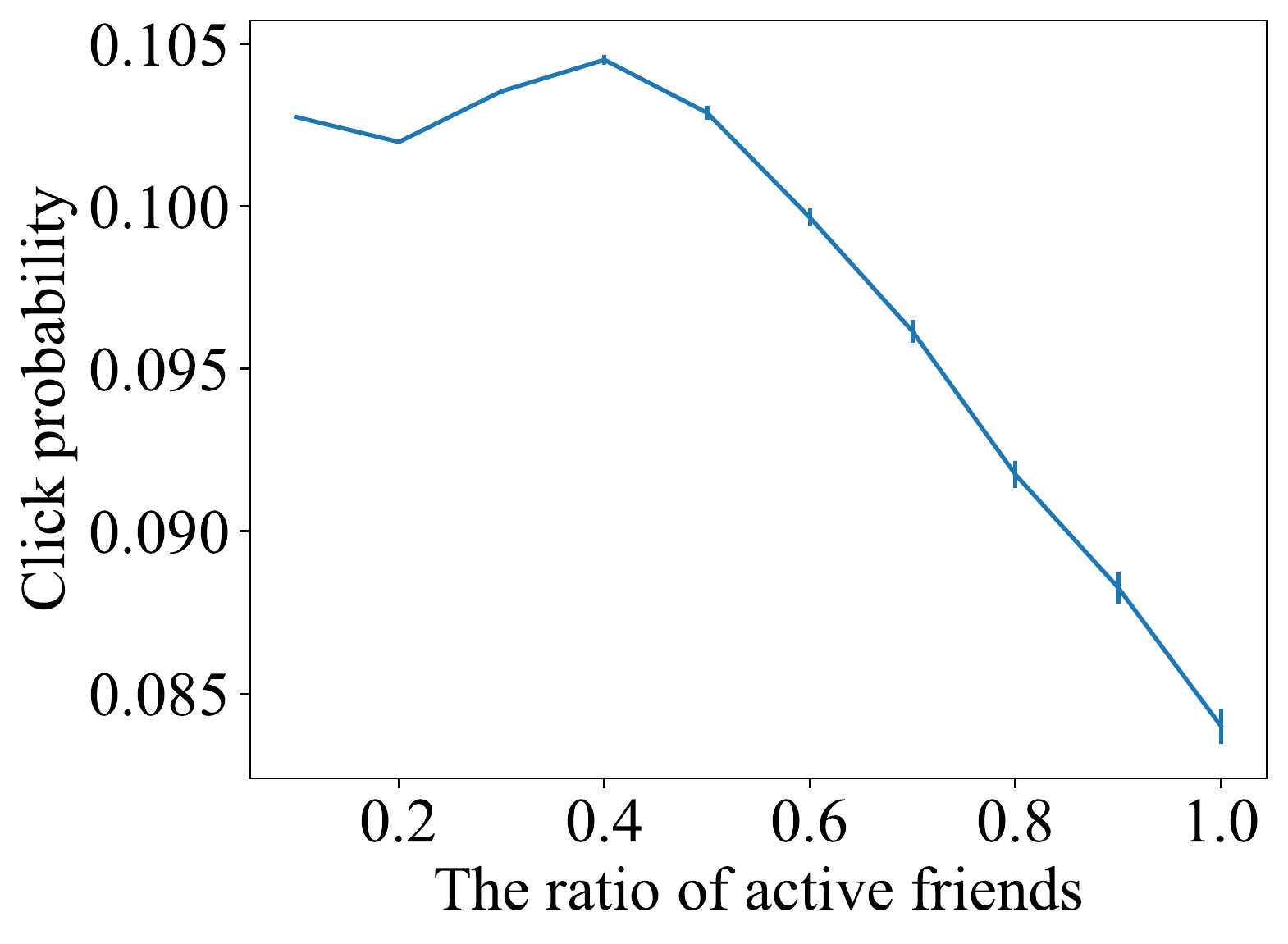}	
			\label{figsub:user-triangle-poi-click}
		}
	}
	\vspace{-0.13in}
	\caption{
		User ``wow'' and click probability vs. the ratio of active friends
	}
	\label{fig:influencefriend}
\end{figure}

\vpara{The Number of Friends in the Ego Networks.} 
Figure \ref{fig:influencefriend} shows how a user’s ``wow'' and click probability on an article changes when the number of active friends increases.
We define the ratio of active friends by dividing a pre-defined maximum number of friends into the actual number of friends.
It demonstrates two very different patterns w.r.t. the two behaviors.
For ``wow'' behavior, with the number of active friends increasing, the probability that the user ``wow''s an article also increases roughly linearly, while for click behavior, the probability first fluctuates a little, and then decreases clearly after the ratio of one's active friends increases to 0.4.
The phenomenon could be explained by information overload --- when the number of one’s
active friends is large, the user may have many other channels from these friends to learn about the information, 
such as ``Moments'' or ``Subscriptions''
(the first is reading articles posted by friends and the other is reading articles of subscribed accounts),
resulting in loss of interest in clicking it~\cite{feng2015competing}.
For ``wow'' behavior, the pattern is consistent with our intuition that 
people may continue to share the hot spots to let more and more people care about them.

\begin{figure*}[t]
	\centering
	\hspace*{0cm}
	\hspace{-0.05in}
	\mbox{
		\subfigure[Users' Active Rate v.s. \#CC in the Ego network formed by active friends]{
			\includegraphics[width=9cm]{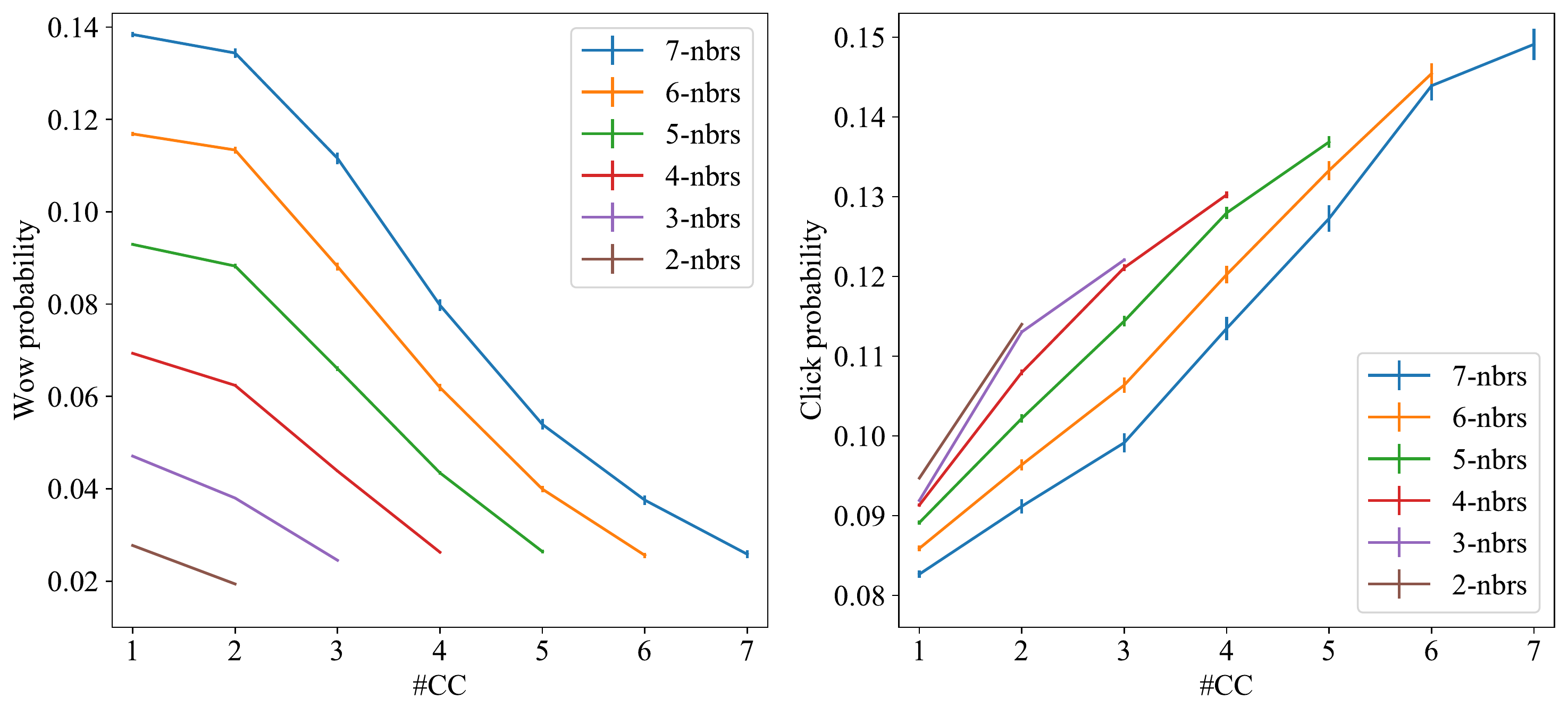}	
			\label{figsub:influencecircle}
		}
		\hspace{0.1in}
		\subfigure[Users' Active Rate v.s. \#CC in the 1-core subgraph of the ego network formed by active friends]{
			\includegraphics[width=9cm]{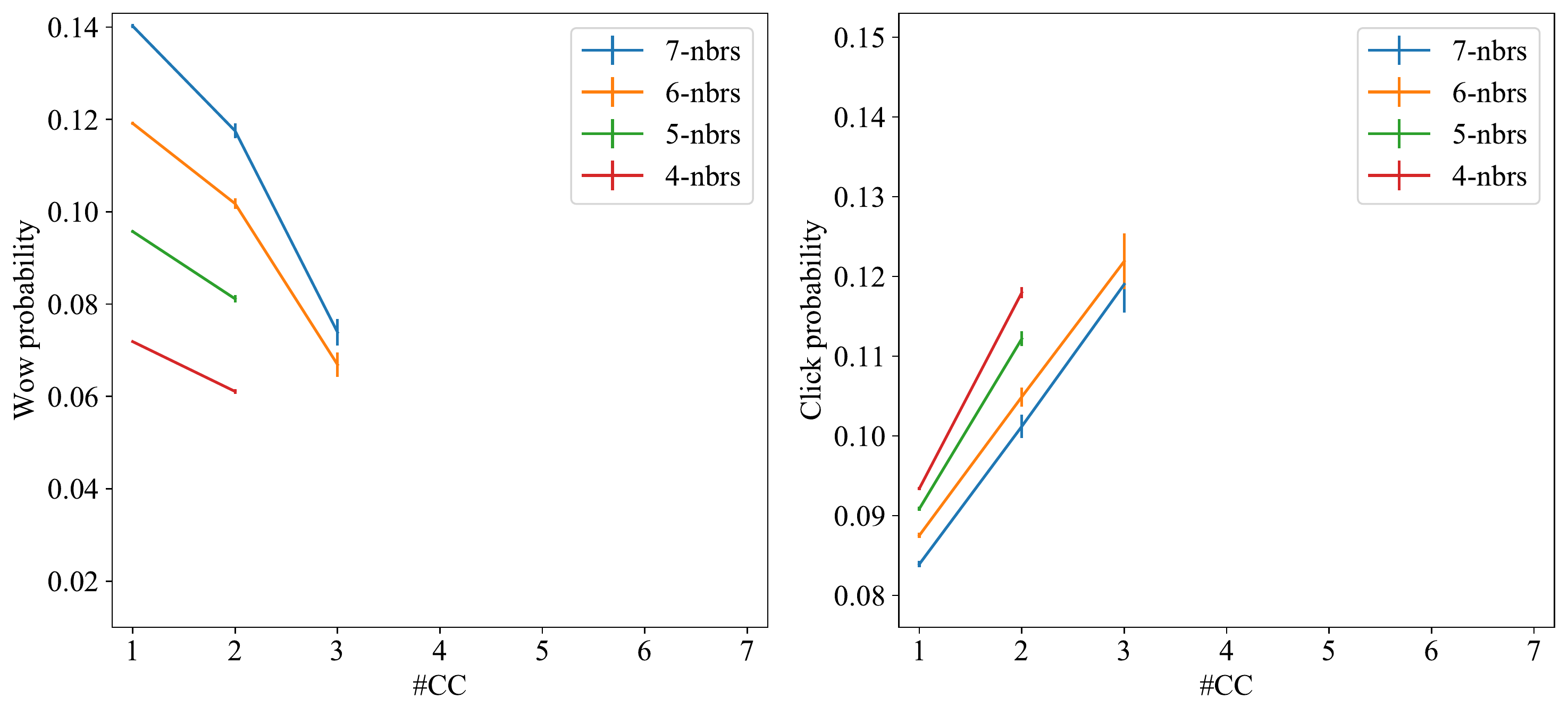}	
			\label{figsub:net-k-core-cc-act}
		}
	}
	\vspace{-0.13in}
	\caption{Social influence analysis: the probability that the user ``wow''s or clicks an article conditioned on 
		the number of connected components of (cleaned) ego networks formed by active friends.
	}
	\label{fig:net_n_friends_like_activity}
\end{figure*}

\vpara{The Number of Connected Components (\#CC) in the Ego Network.} 
Following the above, we conduct another deep analysis, named structural diversity~\cite{ugander2012structural}, to study how the topological structure of one’s active friends would influence the user’s behavior.
Figure \ref{figsub:influencecircle} plots the probability of a user’s behavior w.r.t. the number
of connected components (\#CC) of her/his active friends. Here, each connected component can be viewed as a specific group of friends (connected by friendships among them). 
The pattern is very interesting. When the total number of active friends on an article increases, users are more likely to 
spread the article (see Figure \ref{fig:influencefriend}).
However, when fixing the number of active friends, ranging from 2 to 7, the probability decreases
with the increase of the number of connected components (\#CC) (see Figure \ref{figsub:influencecircle}).
This confirms the structural diversity analysis in
sociology~\cite{ugander2012structural,zhang2013social}, 
which suggests that the user’s interest in
sharing a piece of information will decrease when she/he notices that the information has already been shared by multiple different groups of friends,
since there isn't much benefit for growing the user's influence when many people have shared it.
For click, it is totally different — when a user
notices that multiple different groups of his/her friends have read an article, her/his probability of reading the articles will quickly increase.
If many friends from different circles have ``wow''ed one article,
the article is probably of high quality and has a broad audience, so the user is attracted to read it.

\vpara{\#CC in the Cleaned Ego Network.} 
Although one can have many active friends who ``wow''ed an article, different friends may influence the ego user to a different extent. For example, it is possible that friend $O$ and the ego user became friends by chance, but they are not familiar with each other. Thus, $O$ is an outlier in the ego network who doesn't connect to other friends of the ego user. 
In this case, including friend $O$ in the ego network may introduce noise. 
Thus, we want to first obtain the cleaned ego network, and then analyze the correlation between its structure and the ego user's activity. To obtain the cleaned ego network, we extract the 1-core subgraph of the ego network formed by active friends, where 1-core means the node in the subgraph needs to have at least one edge with other nodes. In this way, outliers are removed from the ego networks.
Figure \ref{figsub:net-k-core-cc-act} plots the probability of a user's behavior w.r.t. \#CC of one's 1-core subgraph of the ego network formed by active friends. Note that 1-core ensures the connectivity of nodes in the subgraph, so some combinations of (\#Friends, \#CC) pairs don't exist, such as 7 friends with 7 CC.
Comparing Figure \ref{figsub:net-k-core-cc-act} with Figure \ref{figsub:influencecircle}, we can see that when fixing the number of active friends (such as 7),  the speed of increase or decrease of ``wow''/click probability w.r.t. \#CC in Figure \ref{figsub:net-k-core-cc-act} is obviously faster.
Such a difference shows that 
the structural topology
of cleaned ego networks probably gives a better discriminative ability to predict ego users' activity.

\subsection{Summaries}
From the above analysis, we have the following discoveries:
\begin{itemize}[leftmargin=*]
	\item \textit{Males are more likely to click but less likely to ``wow'' articles than females.
		Counterintuitively, the young generations (people in their 20s and 30s) have the lowest active rate in \service. }
	\item \textit{For dyadic or triadic correlations, there exists interest homophily between users and friends (such as about gender), but attribute diversity (such as region) also positively correlates with users' activity when there is more than one active friend.}
	\item \textit{According to ego network topology, the patterns of ``wow'' and click behavior are very different. For instance, when fixing the number of active friends, users' ``wow'' probability is negatively correlated to \#CC formed by active friends, but for click behavior, it is the opposite. The patterns can be more significant when the ego network is cleaned.}
\end{itemize}

\section{Predictive Model}

Can we leverage the discovered patterns to predict users' online behaviors? 
In this section, we first briefly formulate the problem and then present our prediction framework.

\subsection{Problem Formulation}

Let $G_u^{\tau} = \{V_u^{\tau}, E_u^{\tau}\}$ be user $u$' s $\tau$-ego network where \textbf{$\boldsymbol{\tau}$-ego network} is a subgraph induced by $u$ and $u$'s $\tau$-degree friends, $V_u^{\tau}$ is the node set of the subgraph $G_u^{\tau}$ 
and $E_u^{\tau}$ is the edge set of $G_u^{\tau}$.
The attribute matrix of users in $V_u^{\tau}$ is denoted as $C_u^{\tau}$. 
When user $u$ is displayed with an article $d$ ``wow''ed (shared) by some friends at timestamp $ts$, we denote 
an action status of user $u$'s ego-network as 
$S_{(u, d, ts)} = \{s_{(v, d, ts)}\in\{0, 1\} | v \in V_u^{\tau} \setminus\{u\}\}$,
where $s_{(v, d, ts)}$ is the action status of user $v$ w.r.t. article $d$ before timestamp $ts$, here 0 means inactive and 1 means active (both denoting ``wow'' behavior).
Our goal is to quantify the ``wow'' and click probability of ego user $u$ after timestamp $ts$:

\beq{
	P(s_{(u, d, >ts)} | G_u^{\tau}, S_{(u, d, ts)}, C_u^{\tau})
}

Since we analyze two different behaviors (click and ``wow'') of users, a straightforward idea is 
to leverage the correlation between click and ``wow'' to design a joint prediction model (like multi-task learning). However, 
there is no evident correlation between click  and ``wow'' in our training set. According to our statistics, $P(is\_click_u = 1)\approx P(is\_click_u=1 | is\_wow_u=1)$, which means that the two behaviors 
are almost independent.
Thus, we choose to learn independent models for predicting the two behaviors.
In the following, we will illustrate our model framework in detail.

\subsection{The \model Framework}

\begin{figure*}[t]
	\centering
	\hspace*{0cm}
	\hspace{-0.05in}
	\includegraphics[width=16cm]{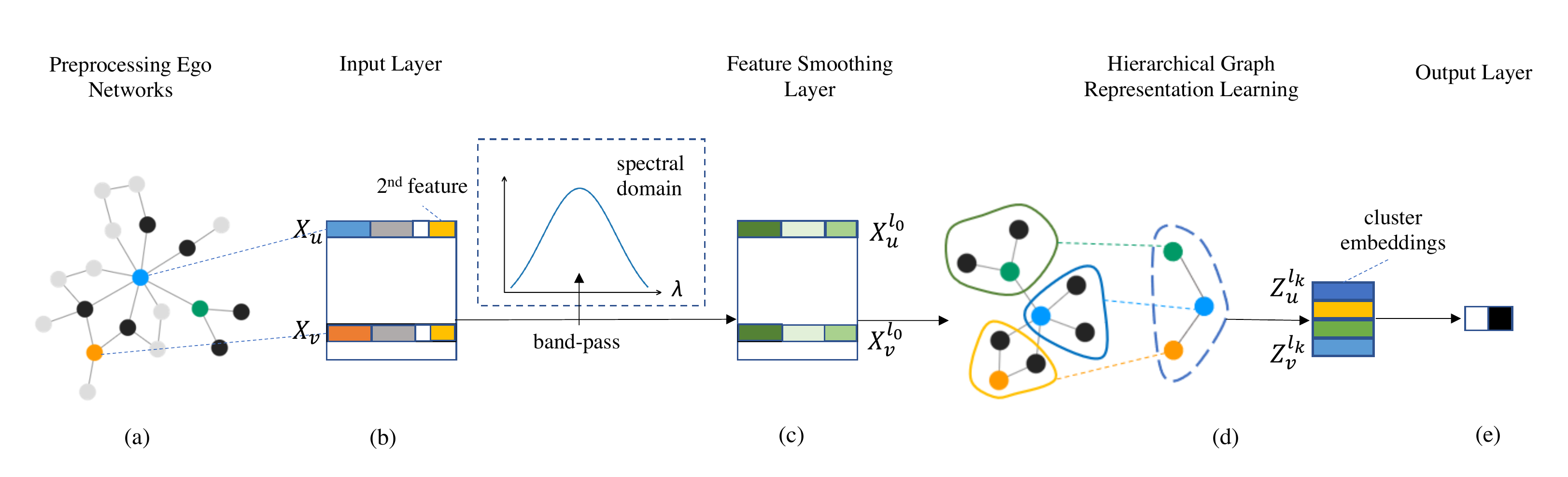}	
	\vspace{-0.2in}
	\caption{Model Framework: \textmd{(a) A sketch map of the processed ego network; 
			(b) The input layer, in which each user's pretrained embedding is concatenated with handcrafted features and her influence feature, which indicates her active status and whether she is an ego user; (c) The feature smoothing layer, in which each user’s concatenated input features are filtered by a band-pass filter in the spectral domain; (d) Each ego network’s  features are passed into a hierarchical graph representation learning model;
			(e) The output layer, where MLP layers are employed to predict user behaviors based on graph representations.}}
	\label{fig:model-framework}
\end{figure*}


In this subsection, we present our proposed model framework \model as illustrated in Figure \ref{fig:model-framework}.
The core model components and the basic idea are as follows:
(1) For input user features, we consider various user features such as user demographics (gender and age) and pre-trained user embeddings, and try to model feature interactions as the analysis in Section \ref{subsec:user_demo}.
(2) We then learn user embeddings by propagating initial features in a trainable modulated spectral domain, by which the learned user embeddings can capture useful information in ego networks and filter out noise, which is motivated by the analysis in Section \ref{sec:ego-net-property}. 
(3) Next, we further feed the learned intermediate representations to a hierarchical graph representation model. 
This model can learn subgraph embeddings by clustering nodes iteratively (here subgraphs can be considered to correspond to connected components analyzed in Sec. \ref{sec:ego-net-property}).
(4) Also, we try to model the interactions between users' features and friends' features with a new attention model as the analysis in Section \ref{subsec:user-relation}.
The proposed \model framework consists of five steps: Preprocessing ego networks, Input layer, Feature smoothing layer, Hierarchical graph representation learning and Output layer.

\vpara{Preprocessing Ego Networks.}
As the $\tau$-ego network can be very large in such a dense social network, especially for users with large degrees, we 
adopt a sampling strategy to sample a subset of users from one's ego-network. 
In this work, we use Breadth-First Search (BFS) to generate a fixed-size ego network for each user/interaction due to its effectiveness and efficiency. 
In detail, we first add the ego user and active friends into the ego network in order. 
The added order of active friends is determined by their active (``wow'') timestamp. 
Then the friends of users in the current ego networks are added by performing BFS iteratively.
Note that Qiu et al.~\cite{qiu2018deepinf} use Random Walk with Restart (RWR)~\cite{haveliwala2003topic} to generate sampled ego networks. 
However, information diffusion in WeChat is very localized (users can only see articles their friends ``wow''ed); 
thus, BFS is more suitable than RWR in this case. Finally, we perform BFS in users' 2-ego networks ($\tau = 2$).
We set the number of nodes in each sampled ego network to $m$. The adjacency matrix of the generated ego network for each instance is denoted as $A_{(u, d, ts)}$ (here an instance refers to an interaction between user $u$ and article $d$ before timestamp $ts$). We omit subscripts $(u, d, ts)$ in the following description if
there is no ambiguity.

\begin{table}
	\caption{List of input features of \model.
	(*x) indicates the dimension of the feature. OL: opinion leader. SH: structural hole.}
	\label{tb:input_features}
	\centering
	\begin{tabular}{c|c|c}
\toprule
		Type & Description & Feature Definition \\
		\hline
		\multirow{3}{*}{Demographics}
		& Gender & (*1) 0: unknown,  1: male, 2: female\\
		&Age & (*1) Age number\\
		& Region & (*10) Encoding via region partition \\
		\hline
		\multirow{2}{*}{Social Roles} 
		& OL & (*1) PageRank score \\
		& SH & (*1) Cut point or not\\
		\hline
		\multirow{2}{*}{Context} 
		& Ego & (*1) Ego user or not \\
		& Action & (*1) ``wow'' or not \\
		\hline
		Embedding & Pre-train &  ProNE embedding \\
		\hline
		Cross Feature & 2nd feature &  FM w.r.t. various features\\
		\bottomrule
	\end{tabular}
\end{table}

\vpara{Input Layer.}
We consider various types of input features for each user, as listed in Table \ref{tb:input_features}.
First, the input layer covers customized user features, such as demographic and social role features.
Second, we also consider two-dimensional contextual features for each user to indicate active status and positions in the ego network, in which one is whether the user ``wow''ed the corresponding article, and the other is whether the user is an ego user~\cite{qiu2018deepinf}. 
Third, we consider pre-trained user embeddings.
Although we study user behavior prediction in the ego network, it would be beneficial to capture user structure information in the global social network.
Thus, we first pre-train user embeddings in the friendship network.
Many network representation learning methods have been proposed to learn node representations in a graph~\cite{perozzi2014deepwalk,tang2015line,zhang2019prone}. 
We adopt ProNE~\cite{zhang2019prone} to pre-train user embeddings in the friendship network, due to its high efficiency and effectiveness, which could take a relatively short time to generate node embeddings of billion-scale graphs, and is effective in capturing global information by propagating embeddings in the spectrally modulated domain.

The above mentioned features can be regarded as first-order user features.
Motivated by the analysis in Subsection \ref{subsec:user_demo},
the cross-attribute factor might also take effect.
Thus, we adopt the factorization machine technique to model feature interactions.
We generate second-order features by first mapping different features into the same space,
and then calculate the second-order feature interactions as follows:

\beq{
	X^{\text{2nd}} = \frac{1}{2}((\sum_{i=1}^{F} W_{i} x^{(i)})^{2} - \sum_{i=1}^{F} (W_{i} x^{(i)})^2 )
}

\noindent where $x^{(i)}$ is $i$-th user feature, $W_{i}$ is the feature projection matrix of feature $x^{(i)}$ and $F$ is the number of different features.
Here the cross-terms indicate the interactions between different features.
Finally, we concatenate all the first-order features $\{x^{(i)}\}_{i=1}^{F}$ and the second-order feature $X^{\text{2nd}}$ to form input feature $X$.

\hide{
The input layer consists of pretrained user embeddings and other features.
Although we study user behavior prediction in the ego network, it would be beneficial to capture user structure information in the global social network.
Thus, we first pre-train user embeddings in the large-scale friendship network.
Many network representation learning methods~\cite{perozzi2014deepwalk,tang2015line,zhang2019prone} have been proposed to learn node representations in a graph. 
We adopt ProNE~\cite{zhang2019prone} to pre-train user embeddings in the large-scale friendship network due to its high efficiency and effectiveness, which could take days to generate node embeddings of billion-scale graphs and is effective by capturing global information by propagating embeddings in the spectrally modulated domain.
Apart from user embeddings, we also consider two-dimensional contextual features for each user to indicate the active status and positions in the ego network, in which one is whether the user ``wow''ed the corresponding article, and the other is whether the user is the ego user~\cite{qiu2018deepinf}. 
Furthermore, the input layer also covers customized vertex features, such as demographic and social role features.
Features of the three parts are concatenated as the input features of users in each ego network.
The features we used are listed in Table \ref{tb:input_features}.
}

\vpara{Feature Smoothing Layer.} 
Pre-trained user embeddings only capture the global network structure. Users residing in different ego networks may play different roles. Thus, they should have different representations in different ego networks. We propose a feature smoothing method via graph filters, which can fine-tune user embeddings $X$ via cleaned ego network structures. 
The output of this step is $X^{l_0}$ by propagating $X$ in the modulated spectral domain of ego networks.

In graph theory, the random walk normalized graph Laplacian is defined as $\mathcal{L} = I_m - D^{-1}A$, where $A$ is the adjacency matrix of the ego network, $m$ is the size of each ego network, $I_m$ is the identity matrix, and $D=\sum_{j} A_{ij}$. The normalized Laplacian can be decomposed as $\mathcal{L} = U\Lambda U^\top$, where $\Lambda=\mathrm{diag}[\lambda_{1}, \lambda_{2},...\lambda_{m}]$. 
In spectral graph theory, small (large) eigenvalues in a graph Laplacian control the network's global clustering (local smoothing) effect, which motivates us to capture useful information of ego networks in the spectral domain.
The global clustering (local smoothing) effect means how well a graph can be partitioned into a small (large) number of clusters, 
so that nodes in different clusters are less connected, while nodes in the same clusters are densely connected.
The smaller the $j$-th eigenvalue $\lambda_{j}$ is, the better partition effect it would achieve for dividing into $j$ clusters~\cite{zhang2019prone}.
Thus we employ a graph filter $g$ to adjust the eigenvalues of the Laplacian.

\beq{
	\widetilde{\mathcal{L}} = U \mathrm{diag} ([g(\lambda_{1}), g(\lambda_{2}), ..., g(\lambda_{m})]) U^\top
}

\noindent where $\widetilde{\mathcal{L}}$ is the modulated Laplacian and $g$ is the spectral modulator. 
We propagate the initial node embeddings in the spectral domain via modulated Laplacian as follows:

\beq{
	X^{l_0} = D^{-1} A ( I_m - \widetilde{\mathcal{L}})X
}

\noindent where $X^{l_0}$ is the user embedding matrix modulated in the spectral domain. Here $I_m - \widetilde{\mathcal{L}}$ is the spectral modulator of the normalized adjacency matrix $D^{-1} A$.
 In this paper, we adopt the graph filter as follows:

\beq{
	g(\lambda)=e^{-\frac{1}{2}[(\lambda-\mu)^2-1]\theta}
}

\noindent where $g$ can be considered an adjustable band-pass filter kernel (see Figure \ref{fig:model-framework}) with $\mu \in [0, 2]$. It passes or enlarges eigenvalues within a certain range and filters out other values, thus reducing the noise or redundant information. 

To avoid explicit eigendecomposition and Fourier transform, we use the same trick as in~\cite{zhang2019prone} to approximate $g$ with a Chebyshev expansion and Bessel function~\cite{andrews1998special}. In our model, we set $\mu$ as a trainable parameter. Thus it can be adaptively learned for different datasets. 

\vpara{Hierarchical Graph Representation Learning.}
As we analyze above, the ego user's activity is strongly correlated to the number of connected components of her ego-network formed by active friends (neighbors). 
Our idea here is to design a hierarchical representation learning method
to encode the substructures of ego networks.
%
Substructures, such as connected components, can be regarded as high-level structural patterns, which motivates us to cluster similar nodes iteratively to encode these substructures.
The goal of this step is to generate high-level ego network representation $Z^{l_k}$ at iteration $k$.
We employ graph neural networks (GNN) ~\cite{niepert2016learning,ying2018hierarchical,sun2019infograph} as basic modules to learn graph representation. 

In detail, we first generate node embeddings of the entire ego networks via a GNN~\cite{kipf2017semi,velivckovic2018graph} module. The input user embeddings are from the learned user embeddings via graph filters described above.

\beq{
	Z^{l_1} = \text{GNN}_{0, \text{embed}}(A^{l_0}, X^{l_0}) \label{eq:gnn_0_emb}
}

\noindent where $Z^{l_k}$ is the hidden node embedding of layer $l_k$, $A^{l_0}$ is the adjacency matrix of ego networks, and $X^{l_0}$ is the input node features. 
In order to generate coarsened graphs to represent graph substructures, 
following DIFFPOOL~\cite{ying2018hierarchical},
we learn an assignment matrix $B^{l_{k+1}}$ via another GNN,

\beq{
	B^{l_{k+1}} = \text{softmax} (\text{GNN}_{k, \text{pool}} (A^{l_{k}}, X^{l_{k}})) \label{eq:gnn_assign}
}

\noindent where $B^{l_{k+1}} \in R^{m_k \times m_{k+1}} (m_{k+1} < m_k, m_0 = m)$ and $b_{ij}^{l_{k+1}}$ represents the probability of assigning node $i$ to $j$-th clusters in $(k+1)$-th assignment layer .

With assignment matrix $B^{l_k}$, an ego network can be transformed into a smaller graph iteratively, in which each node represents a ``cluster''.

\beq{
	X^{l_k} = {B^{l_k}}^\top Z^{l_k} \in \mathbb{R}^{m_k \times h_k}
}

\beq{
	A^{l_k} = {B^{l_k}}^\top A^{l_{k-1}} B^{l_k} \in \mathbb{R}^{m_k \times m_k}
}

\noindent where $X^{l_k}$ is the cluster embeddings, and $A^{l_k}$ is the coarsened adjacency matrix which denotes the connectivity strength between pairwise clusters.

Based on the coarsened graph, the coarse-level node embeddings can be generated by

\beq{
	Z^{l_{k+1}} = \text{GNN}_{k, \text{embed}} (A^{l_k}, X^{l_k})
}

We generate different levels of (sub)graph embeddings by pooling operations on node embedding matrices, and then concatenate them to form the final representations of ego networks,

\beq{
	Z^{\text{graph}} = \bigparallel\limits_{k=1}^{L} \sigma(Z^{l_k})\label{eq:graph-pool}
}


\noindent We set $\sigma$ in Eq. \ref{eq:graph-pool} as a dimension-wise max-pooling operation or sum-pooling to transform node embedding matrix into graph embedding.

\vpara{\textit{Basic GNN modules.}} 
As for GNN modules in Eq. \ref{eq:gnn_0_emb}, Eq. \ref{eq:gnn_assign} and Eq. \ref{eq:graph-pool},
we argue that GAT~\cite{velivckovic2018graph} is suitable, since it can learn the attention weights of neighboring nodes with their node features.
GAT learns the attention weight between node $i$ and node $j$ as follows:

\beq{
	\alpha_{ij}^{\text{AA}} = \frac{\text{exp}(\text{act}(a^{\top}_{\text{src}} W_p x_i+ a^{\top}_{\text{dst}} W_p x_j))
	}
	{\sum_{t \in \mathcal{N}_i} \text{exp}(\text{act}(a^{\top}_{\text{src}} W_p x_i + a^{\top}_{\text{dst}} W_p x_t))
	}
\label{eq:gat_attn}
}

\noindent where $\mathcal{N}_i$ is the neighbors of node $i$, $x_i$ is the hidden embedding of node $i$, $W_p$ is the feature projection matrix, $a_{\text{src}}$ and $a_{\text{dst}}$ are attention parameter, and $\text{act}$ is the LeakyReLU activation function.
As shown in Eq. \ref{eq:gat_attn}, GAT uses \textit{additive attention (AA)}.

However, the attention mechanism used in GAT doesn't consider feature interactions between neighboring nodes.
Thus, we employ a simple modification over Eq. \ref{eq:gat_attn}:

\beq{
	\alpha_{ij}^{\text{DA}} = \frac{\text{exp}(\text{act}((a^{\top}_{\text{src}} W_p x_i + b_{\text{src}} ) \cdot (a^{\top}_{\text{dst}} W_p x_j + b_{\text{dst}})))
	}
	{\sum_{t \in \mathcal{N}_i} \text{exp}(\text{act}((a^{\top}_{\text{src}} W_p x_i + b_{\text{src}}) \cdot  (a^{\top}_{\text{dst}} W_p x_t + b_{\text{dst}})))
	}
	\label{eq:gdt_attn}
}

\noindent where $b_{\text{src}}$ and $b_{\text{dst}}$ are bias terms.
We term this attention \textit{Dot attention (DA)}.
As shown in Eq. \ref{eq:gdt_attn}, the cross terms model feature interactions of neighboring nodes.
We denote our model using GAT modules as \modelat and the model using dot attention as \modeldt.

\vpara{Output Layer.}
Finally, we let the ego network embedding $Z^{\text{graph}}$ pass into fully connected layers to generate the 
prediction scores, which are used to compare with the ground-truth ``wow''/click labels.
We use the cross-entropy loss as our objective function.
The prediction function at the output layer and the loss function are described in Eq. \ref{eq:pred} and Eq. \ref{eq:loss}, respectively, where $fc_{\text{pred}}$ represents the fully-connected layers,
$y_{\text{pred}}^{i}$ denotes the action probability of instance $i$, and
 $y_i$ is the ground-truth of instance $i$.

\beq{
	y_{\text{pred}} = fc_{\text{pred}}(Z^{graph}) \label{eq:pred}
}
\beq{
	\mathcal{L} = -\sum_{i=1}^{N} (y_i \times \log (y_{\text{pred}}^{i}) + (1 - y_i) \times (1 - \log (y_{\text{pred}}^{i}) ) ) 
	\label{eq:loss}
}

\subsection{Discussion}

When users' click and ``wow'' behaviors are concerned, 
it is natural to take into account the article content.
If we further consider articles, 
the problem is highly related to social recommendation.
However, the main focus of this article is user behavior prediction based on user demographics, correlations between users, and ego network properties,
which is related to the problem of social influence locality, 
so we exclude article information to make the problem clear.

\hide{
Our model framework is closely related to and motivated by the insights from data, which is different from many neural-based methods.
Specifically, (1)
\textbf{To model cross-attribute factor for users' different attributes} (such as user embedding and demographics), we adopt factorization machine technique to generate second-order user features to model feature interactions for each individual.
(2) \textbf{To remove noises in the ego network},  ProHENE propagates initial user features in the modulated spectral domain,
to generate user embeddings based on cleaned ego networks.
(3) \textbf{To model user relationships},
we adopt a new graph attention mechanism to model feature interactions between neighbors.
(4) \textbf{To model the connected components --- hierarchical structure of the ego networks},
we generate hierarchical representations of ego networks by clustering nodes together and learning on coarsened graph iteratively.
}
\section{Experiments}

We present the effectiveness of our model on users' ``wow'' and click behavior prediction of WeChat \service, and also a publicly available Weibo dataset. 
The codes are publicly available.\footnote{\scriptsize{\url{https://github.com/zfjsail/wechat-wow-analysis}}}

\subsection{Experiment Setup}

\begin{table*}
	\newcolumntype{?}{!{\vrule width 1pt}}
	\newcolumntype{C}{>{\centering\arraybackslash}p{2.2em}}
	\caption{
		\label{tb:performance} Results of User Behavior Prediction.
		\normalsize
	}
	\centering\small
	\renewcommand\arraystretch{1.0}
	\begin{tabular}{c|@{~ }*{1}{CCCC|}*{1}{CCCC|}CCCC}
		\toprule
		&\multicolumn{4}{c|}{WeChat ``Wow''}
		&\multicolumn{4}{c|}{WeChat Click} 
		&\multicolumn{4}{c}{Weibo}
		\\
		\cmidrule{2-5} \cmidrule{6-9} \cmidrule{10-13}
		{Method}   & {Prec} & {Rec} & {F1}  & {AUC} & {Prec} & {Rec} & {F1} & {AUC} & {Prec} & {Rec} & {F1} & {AUC}  \\
		\midrule
		Random
		& 47.84 & 50.06 & 48.92 & 50.05
		& 28.64  & 50.13 & 36.45  & 50.02
		& 25.12 & 50.48 &  33.55 & 50.15
		\\
		LR
		& 68.08  & 70.08  & 69.06 & 76.73
		& 41.71  & 67.01  & 51.41  & 70.07
		& 42.97 & 71.37 & 53.64 & 76.38
		\\
		RF~\cite{liaw2002classification}
		& 69.20  & 65.17  & 67.12 & 76.69 
		& 39.52  & 74.69 & 51.69 & 70.12
		& 40.03 & 73.66 & 51.87 & 75.14
		\\
		xDeepFM~\cite{lian2018xdeepfm}
		& 66.23  &  80.96 & 72.85  & 78.25
		& 40.88 & 75.09  & 52.94  & 71.61
		& 30.20 & 73.90 & 42.88 & 64.38
		\\
		DeepInf~\cite{qiu2018deepinf}
		& 70.28  & \textbf{81.46} & 75.46  & 83.06
		& 43.88 & 76.03  & 55.65  & 74.50
		& 48.09 & 71.67 & 57.56 & 81.46
		\\
		Wang et al.~\cite{wang2020social}
		& 69.76 & 79.40 & 74.27 & 81.91
		& 41.91 & 75.07 & 53.79 & 72.31
		& 45.58 & 74.63 & 56.59 & 80.26
		\\
		SAGPool~\cite{lee2019self}
		& 81.74 & 75.43 & 78.46 & 86.18
		& 46.58 & 79.19 & 58.66 & 77.37
		& 43.79 & 73.81 & 54.97 & 78.89
		\\
		ASAP~\cite{ranjan2020asap}
		& 71.13 & 79.81 & 75.22 & 83.28
		& 44.92 & 76.57 & 56.62 & 75.48
		& 46.55 & 70.64 & 56.12 & 79.87
		\\
		StructPool~\cite{yuan2020structpool}
		& 67.56 & 79.21 & 72.92 & 79.46
		& 40.20 & 78.55 & 53.19 & 71.54
		& 30.47 & 72.87 & 42.98 & 61.83
		\\
		\hline
		\modelat
		& 84.95 & 76.81 & \textbf{80.67} & 87.64
		& \textbf{46.63} & 82.01 & 59.46 & 78.05
		& \textbf{50.09} & 72.87 & \textbf{59.37} & \textbf{83.08}
		\\
		\modeldt
		& \textbf{85.46} & 76.30 & 80.62 & \textbf{87.69}
		& 46.45 & \textbf{82.81} & \textbf{59.52} & \textbf{78.27}
		& 48.70 & \textbf{74.88} & 59.01 & 82.76
		\\
		\hline
		\hline
		w$/$o pre-train
		& 74.96 & 78.42 & 76.65 & 84.91
		& 45.68 & 75.77 & 57.00 & 76.09
		& 47.33 & 74.15 & 57.78 & 81.51
		\\
		w$/$o node feature
		& 85.01 & 75.69 & 80.08 & 87.10
		& 45.64 & 82.26 & 58.71 & 77.64
		& 46.34 & 74.46 & 57.13 & 81.10
		\\
		w$/$o 2nd feature
		& 86.40 & 76.16 & 80.16 & 87.50
		& 46.66 & 81.66 & 59.39 & 78.16
		& 46.12 & 75.02 & 57.12 & 81.03
		\\
		w$/$o smoothing
		& 79.23 & 77.57 & 78.39 & 86.04
		& 46.26 & 78.38 & 58.18 & 76.89
		& 48.95 & 72.20 & 58.35 & 82.13
		\\
		\bottomrule
	\end{tabular}
	
\end{table*}

\vpara{Datasets.} 
We mainly evaluate our model on the collected WeChat \service dataset.
To further verify the generalization ability of our method, we also choose a publicly available Weibo dataset for evaluation.

For WeChat \service dataset,
we randomly sample a subset of data to evaluate our proposed method.
To effectively model the influence of users' friend circles on users' online behaviors, we only consider interactions with at least five friends having ``wow''ed the articles. 
After filtering out data, we select all positive instances, in which
there are \num{3163171} ``wow'' instances, \num{2181279} click instances, which result
in \num{5121571} positive instances in total. We also sample a subset of negative instances randomly
to keep the ratio between positive and negative instances relatively balanced.
We divide training/validation/testing sets according the timestamps of the interactions.
That is to say, data from the earlier time are used for training/validation and others for testing.
We keep the ratio between positive and negative instances at about $1.5: 1$ and $1: 1$ for  ``wow'' and click datasets respectively.

Another dataset is a Weibo dataset\footnote{\url{http://aminer.org/Influencelocality}}.
Weibo\footnote{\url{https://weibo.com}} is the most popular microblogging system in China.
The original dataset consists of direct user following networks and tweet logs in 2012.
The goal is to predict users' retweet behavior based on their local neighbors.
We follow the same setup as in~\cite{qiu2018deepinf}.
Finally, we obtain \num{779164} data instances, in which $50\%$ are used for training, $25\%$ for validation, and $25\%$ for testing.

\hide{
\begin{table}[h]
	\centering
	\begin{tabular}{c|c}
		\hline
		Category & Description   \\ \hline
		\multirow{ 2}{*}{Ego User} & 0  \\
		&  \\ \hline
	\end{tabular}
	\caption{A test caption}
	\label{table2}
\end{table}
}

\vpara{Comparison Methods.} We compare our proposed model with the following methods.

\begin{itemize}[leftmargin=*]
	\item \textbf{Random.} We generate like/click probability uniformly in the range [0, 1) for prediction.
	\item \textbf{Logistic Regression (LR).} We use logistic regression (LR) to train a classification model. We define three categories of features: (1) ego users' features, including user gender, age, region, social roles (whether one is an opinion leader or a structural hole) 
	(2) ego network features, including the number of active friends, the number of connected components (\#CC), and the local clustering coefficient of the ego graph formed by active friends; 
	(3) correlation features of ego users and friends: average and sum of the common friends' ratio between the ego user and each active friend. 
	\item \textbf{Random Forest (RF)~\cite{liaw2002classification}.} We use Random Forest to train a classification model due to its effectiveness in selecting relevant features and instances. The features used are the same as Logistic Regression.
	\item \textbf{xDeepFM~\cite{lian2018xdeepfm}.} xDeepFM is a framework based on the Factorization Machine (FM). 
	It learns high-order feature interaction with FM modules 
	and also has DNN (feed forward networks) modules to model feature interactions implicitly.
	The input features are the same as LR and RF.
	\item \textbf{DeepInf~\cite{qiu2018deepinf}.} DeepInf is a framework to learn users' latent representation for predicting social influence. It takes users' ego networks as input, and uses the graph neural network to learn user representation. Here we adopt GAT to learn user embedding due to its superiority for influence prediction in the paper~\cite{qiu2018deepinf}.
	
	\item \textbf{Wang et al.~\cite{wang2020social}.} 
	This method models the topological influence structure based on the Weisfeiler-Lehman (WL) algorithm,
	and learns the influence dynamics for the ego user by leveraging GAT, too.
	The different parts of features are concatenated to make predictions.
	
	\item \textbf{SAGPool~\cite{lee2019self}.}
	SAGPool is a graph pooling method that uses self-attention to distinguish between nodes that should be dropped and the nodes that should be retained.
	The predictions are made on smaller graphs.
	
	\item \textbf{ASAP~\cite{ranjan2020asap}.}
	ASAP utilizes a novel self-attention network to cluster similar nodes together in a graph.
	Then, the most important clusters are selected and included in the pooled graph.
	After each pooling step, the graph is summarized using a readout function.
	
	\item
	\textbf{StructPool~\cite{yuan2020structpool}.}
	StructPool is a hierarchical graph pooling method, which formulates the cluster assignment problem as a structured prediction problem.
	It employs conditional random fields to capture the relationships among assignments of different nodes.
	
	\item \textbf{\model.} Our model takes users' ego networks and user features as input. A delicate graph filter is used to transform user features in the modulated spectral domain of ego networks.
	Then we learn hierarchical structure embeddings of ego networks to predict ego users' behavior in an end-to-end fashion.
	We use \textbf{\modelat} to denote using additive attention in the basic GNN modules, and use \textbf{\modeldt} to denote using dot attention in the basic GNN modules.
\end{itemize}

\vpara{Parameter Settings.}
For the WeChat dataset, we set the maximum number of users/nodes in the sampled ego network to 32. For pre-trained user embeddings, we generate 64-dim embeddings using ProNE~\cite{zhang2019prone}.
In the user feature smoothing method via graph filter, we choose the parameters in the graph filter as follows: $\mu = 0.4$, $\theta = 7$.
For hierarchical graph representation learning, the number of graph coarsening steps is set to 2.
In GAT encoders of hierarchical graph representation learning, we set the head number to 8 and hidden vector dimension for each head to 16.
When training, the learning rate is 0.01 for ``wow'' prediction and 0.1 for click prediction.
The L2 regularization weight is 0.0005. 
Adagrad~\cite{duchi2011adaptive} is chosen as the optimizer.
As for the Weibo dataset, there are several differences as follows.
Following~\cite{qiu2018deepinf}, we adopt random walk with restart (RWR) to generate the sampled ego networks.
The number of graph coarsening steps is 1 and the learning rate is 0.05.

\subsection{Overall Results}

\begin{figure*}[t]
	\centering
	\hspace*{0cm}
	\hspace{-0.05in}
	\mbox{
		
		\subfigure[WeChat ``Wow'']{
			\includegraphics[width=5cm]{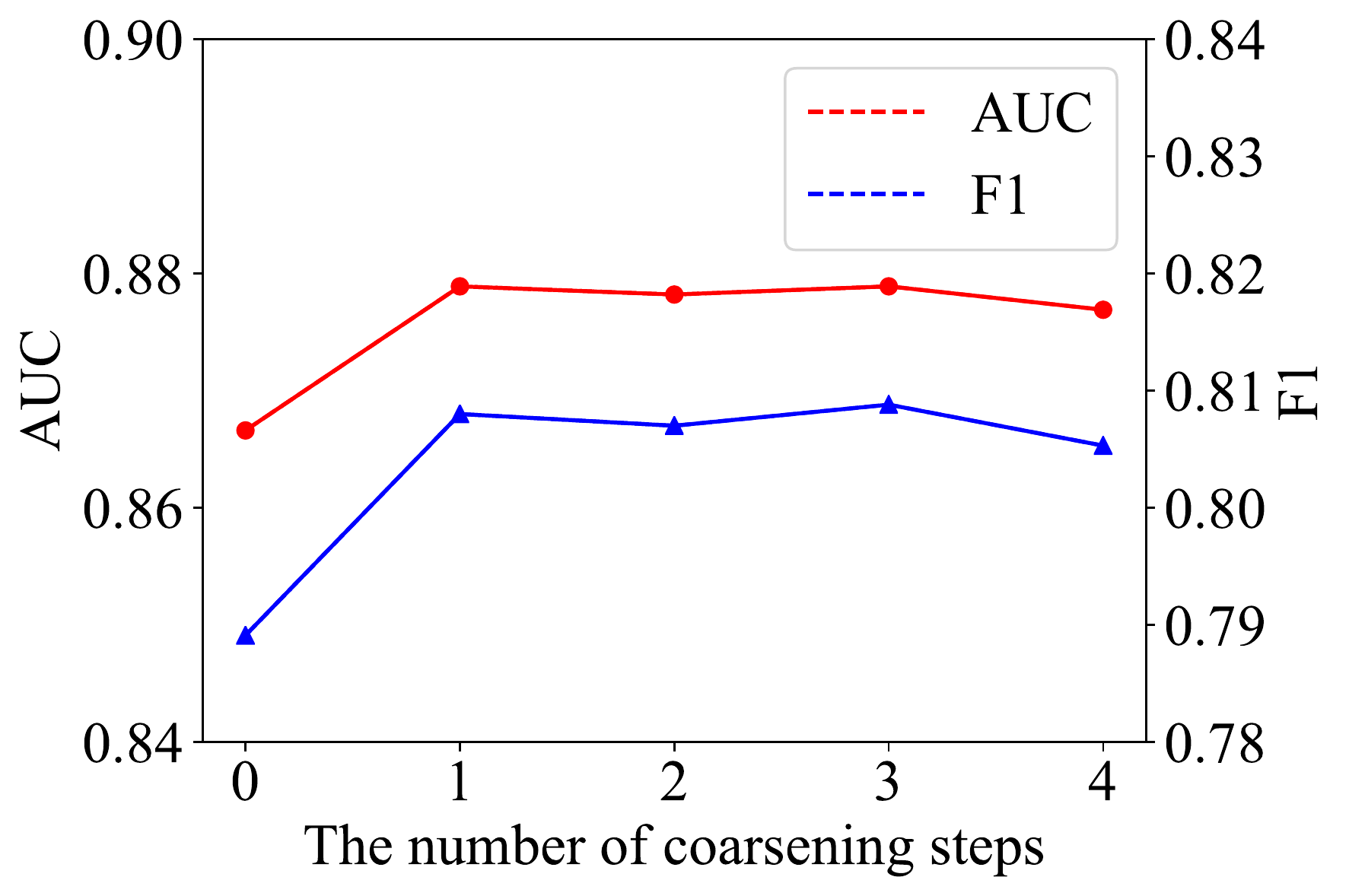}	
			\label{figsub:num-pooling-wow}
		}
		\hspace{0.1in}
		
		\subfigure[WeChat Click]{
			\includegraphics[width=5cm]{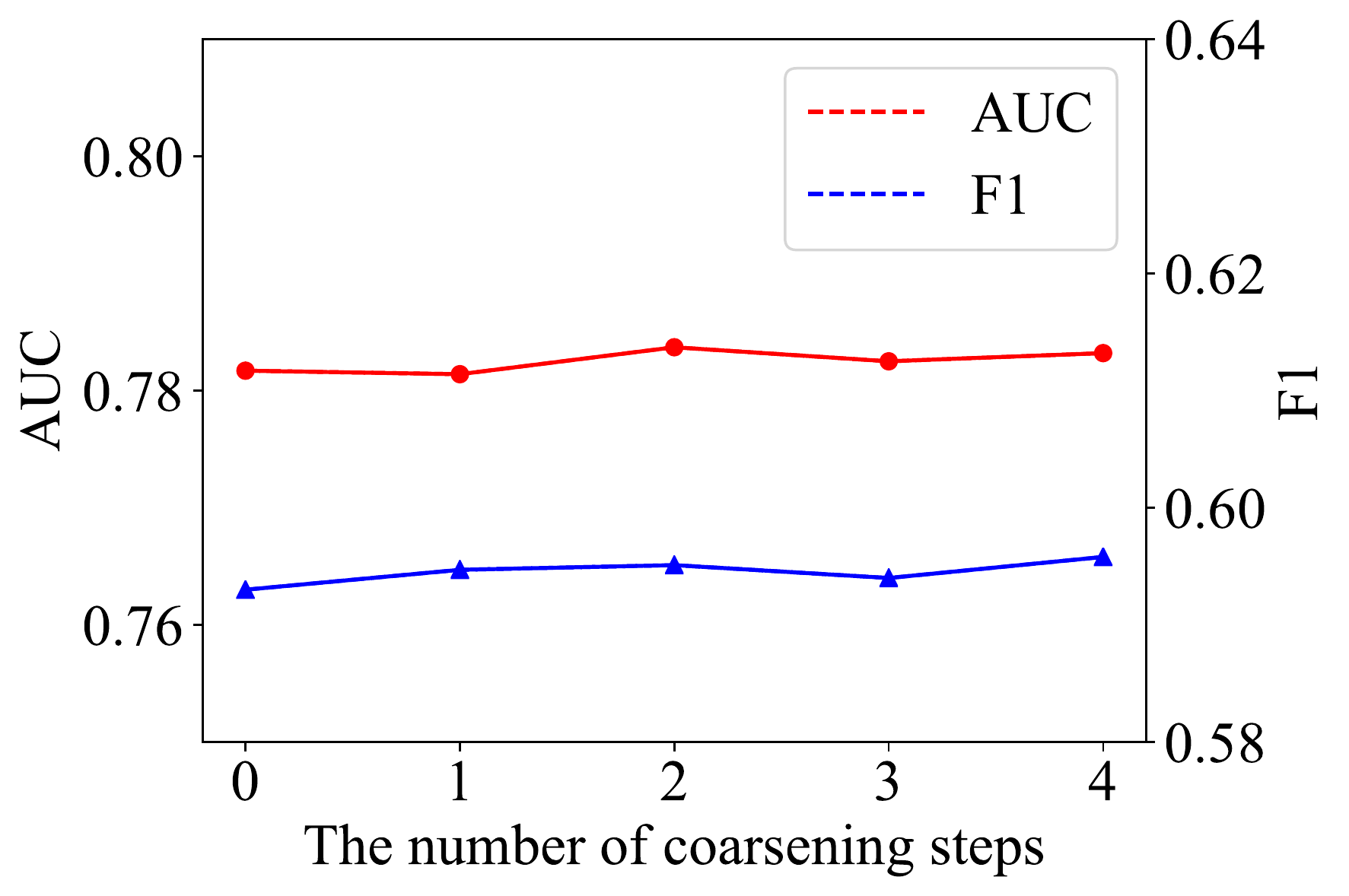}	
			\label{figsub:num-pooling-click}
		}
		\hspace{0.1in}
		\subfigure[Weibo]{
			\includegraphics[width=5cm]{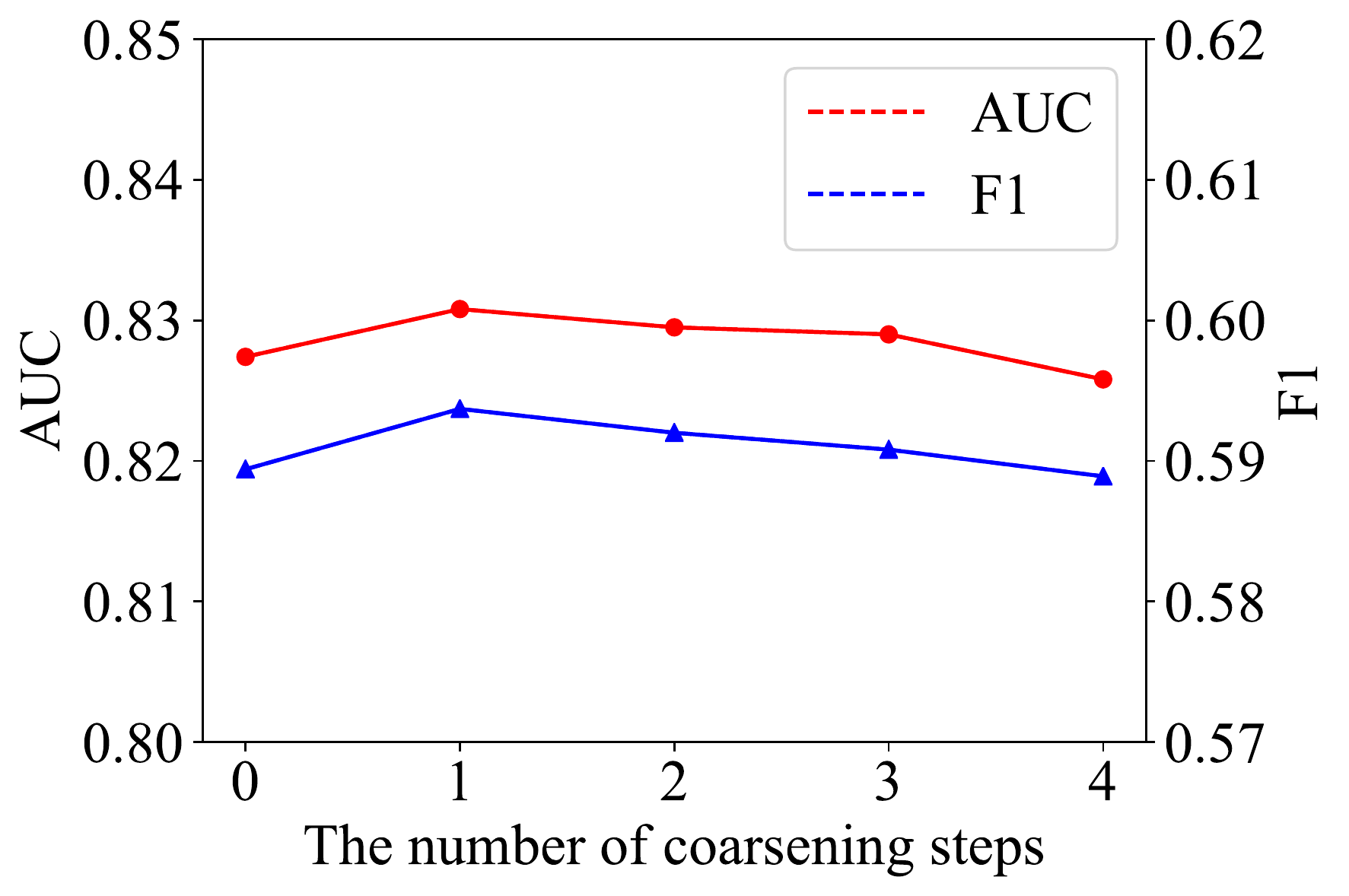}	
			\label{figsub:num-pooling-wow1}
		}
		
	}
	\vspace{-0.13in}
	\caption{``Wow'' and click performance (AUC) on test dataset w.r.t. the number of pooling layers in hierarchical graph representation learning
	}
	\label{fig:num-pooling-effect}
\end{figure*}

\begin{figure*}[t]
	\centering
	\hspace*{0cm}
	\hspace{-0.05in}
	\mbox{
		
		\subfigure[WeChat ``Wow'']{
			\includegraphics[width=5cm]{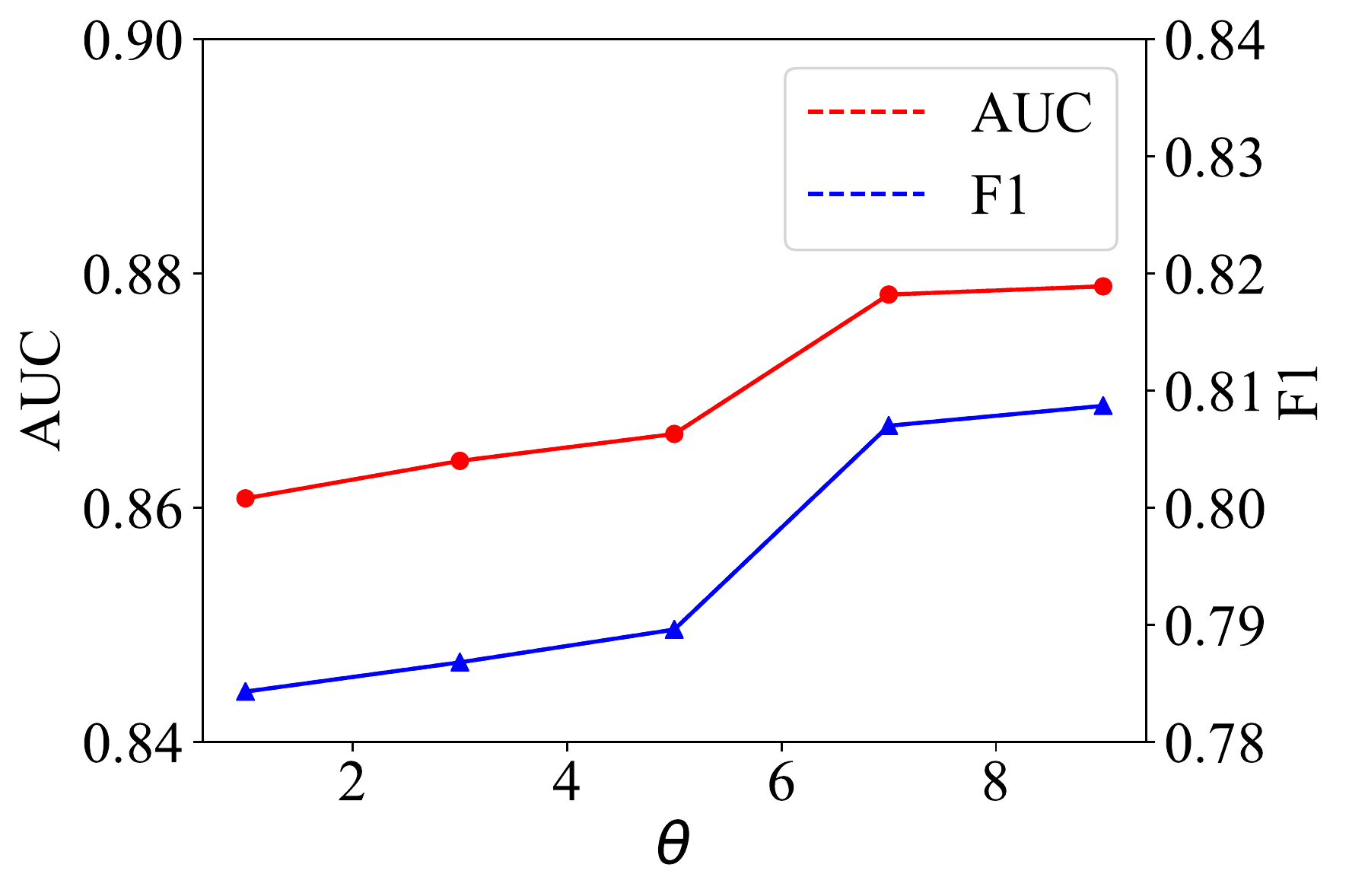}	
			\label{figsub:theta-wow}
		}
		\hspace{0.1in}
		
		\subfigure[WeChat Click]{
			\includegraphics[width=5cm]{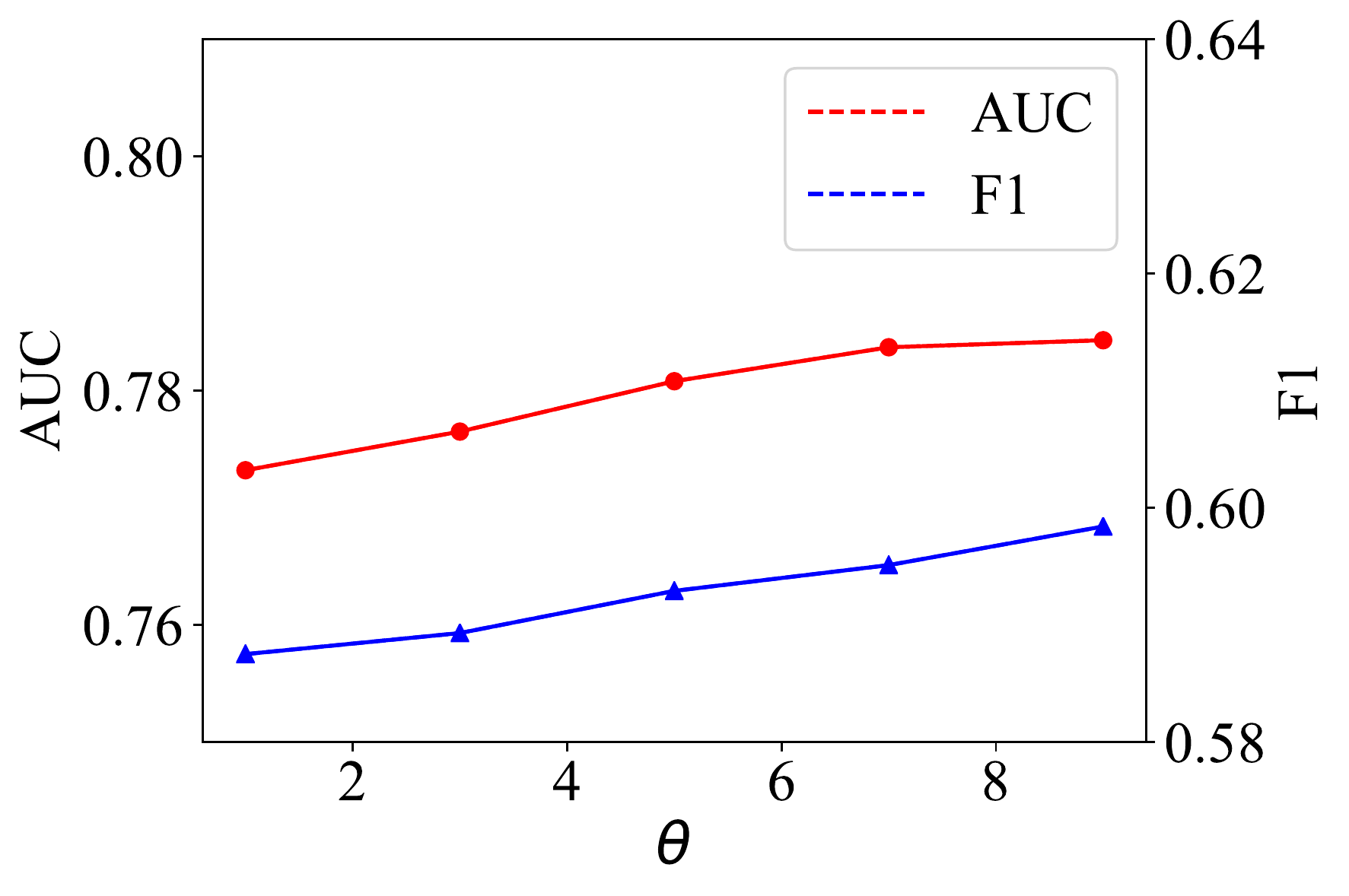}	
			\label{figsub:theta-click}
		}
		\hspace{0.1in}
		\subfigure[Weibo]{
			\includegraphics[width=5cm]{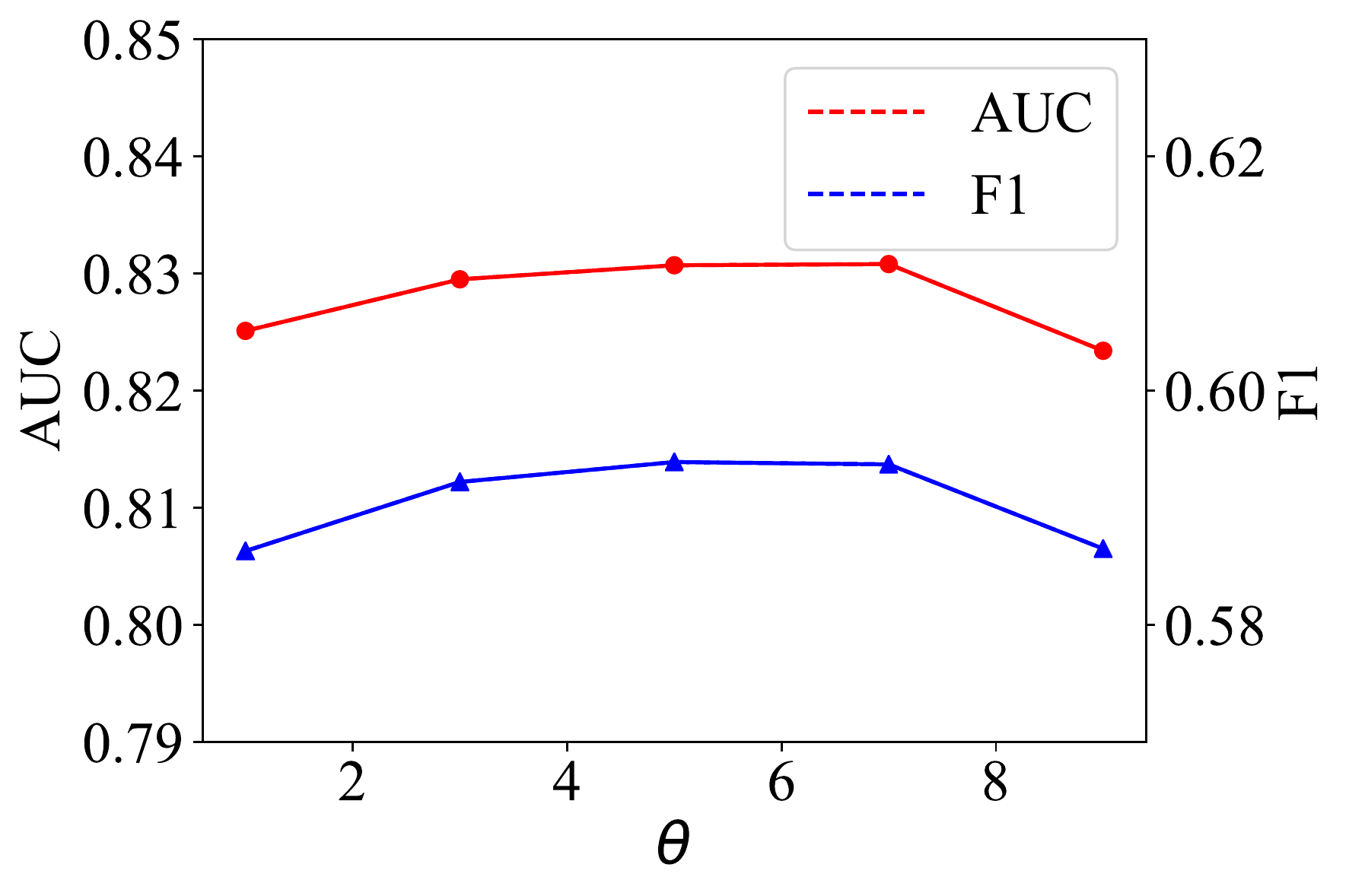}	
			\label{figsub:theta-weibo}
		}
		
	}
	\vspace{-0.13in}
	\caption{
		Prediction performance (AUC and F1) on test dataset w.r.t. $\theta$ in the graph filter of the feature smoothing step.
	}
	\label{fig:theta-performance}
\end{figure*}

\hide{
\begin{table}
	\caption{Wow prediction performance.}
	\label{tb:like-prediction}
	\centering
	\begin{tabular}{c|c|c|c|c}
		\hline
		\textbf{Method} & \textbf{Precision} & \textbf{Recall} & \textbf{F1} & \textbf{AUC} \\
		\hline
		Random & 47.84 & 50.06 & 48.92 & 50.05 \\
		\hline
		LR & 68.08  & 70.08  & 69.06 & 76.73  \\
		\hline
		RF & 69.20  & 65.17  & 67.12 & 76.69  \\
		\hline
		xDeepFM & 66.23  &  80.96 & 72.85  & 78.25 \\
		\hline
		DeepInf & 70.28  & \textbf{81.46} & 75.46  & 83.06 \\
		\hline
		\model & \textbf{84.17} & 76.64 & \textbf{80.23}  & \textbf{87.26} \\
		\hline
	\end{tabular}
\end{table}
}

\hide{
\begin{table}
	\caption{Click prediction performance.}
	\label{tb:click-prection}
	\centering
	\begin{tabular}{c|c|c|c|c}
		\hline
		\textbf{Method} & \textbf{Precision} & \textbf{Recall} & \textbf{F1} & \textbf{AUC} \\
		\hline
		Random & 28.64  & 50.13 & 36.45  & 50.02 \\
		\hline
		LR  & 41.71  & 67.01  & 51.41  & 70.07 \\
		\hline
		RF & 39.52  & 74.69 & 51.69 & 70.12  \\
		\hline
		xDeepFM & 40.88 & 75.09  & 52.94  & 71.61 \\
		\hline
		DeepInf & 43.88 & 76.03  & 55.65  & 74.50 \\
		\hline
		\model & \textbf{46.50}  &  \textbf{81.24} & \textbf{59.14}  & \textbf{77.90}  \\
		\hline
	\end{tabular}
\end{table}
}

Table \ref{tb:performance} summarizes the results of user behavior prediction.
The comparison methods can be roughly divided into several categories: 
(1) traditional classifiers: LR and RF, 
(2) deep learning method by modeling feature interactions: xDeepFM, 
(3) the state-of-the-art user behavior prediction methods based on ego networks: DeepInf and Wang et al.
and (4) hierarchical graph representation learning methods: SAGPool, ASAP and StructPool.
(3) and (4) are both GNN-based methods.
Generally, we observe that our model \model consistently outperforms baseline methods.

For traditional classifiers, it can not achieve better prediction performance than other methods, 
although it leverages hand-craft user features, correlation features, and network features.
xDeepFM, a factorization-machine based neural network model, achieves better performance than LR and RF on WeChat dataset, which might imply that the correlation between user features is an inherent factor that impacts users' ``wow'' and click behaviors, such as the correlation between users' gender and age.

DeepInf and Wang et al. could both achieve good prediction performance on three datasets.
It demonstrates that modeling dyadic correlations via graph attention is effective.
However, the prediction performance of Wang et al. is inferior to DeepInf,
which might indicate that sometimes local topological features could result in a negative impact on user behavior prediction.

For hierarchical graph representation learning methods, SAGPool outperforms most of the baselines, though still weakly than \model. 
This indicates that dropping nodes via self-attention on graphs is another effective solution to graph coarsening.
Moreover, ASAP performs better than SAGPool on the Weibo dataset, 
which might imply that Weibo and WeChat datasets have different characteristics for user behavior modeling.
As for StructPool, we infer that modeling cluster assignment as a structured prediction problem via conditional random fields is ineffective for our problem, due to its inferior performance. 

For our proposed model, its superiority is mainly due to the hierarchical structure embedding of ego networks and the feature smoothing effect.
Moreover, compared with baselines using various hand-crafted features, 
our method only uses user features as input. 
Thus, our method can model dyadic correlations and ego network structures better than hand-crafted features.
Furthermore, 
we observe that \modeldt achieves similar performance as \modelat and
 \modeldt sometimes outperforms \modelat slightly.
Here we argue that AUC is more important than F1 metric, since F1 depends on a good threshold for classification.
Thus, modeling feature interactions between neighboring nodes takes effect at times.

In addition, for the WeChat \service dataset, we find that the performance differences between different methods for click prediction are smaller than those of ``wow'' prediction.
Another observation is that, in general, the click prediction performance is much lower than ``wow'' prediction performance. 
This phenomenon is probably because users' click (or reading) behavior is more correlated to the articles themselves, while their ``wow'' behavior is more relevant to the social influence around them.

\subsection{Ablation Study}
We study the effects of different model components for user behavior prediction.
\begin{itemize}[leftmargin=*]
	\item \textbf{w$/$o pre-train}: Remove the pre-trained ProNE user embedding in the input features. Note that the second-order features also lack this part.
	\item \textbf{w$/$o node feature}: Remove the demographic and social role features in the input features. 
	Note that the second-order features also lack this part.
	\item \textbf{w$/$o 2nd feature}: Remove the second-order feature interactions of different user features in the input features.
	\item \textbf{w$/$o smoothing}: Remove the feature smoothing step via the graph filter.
\end{itemize}

In Table \ref{tb:performance}, the bottom part summarizes the results of the ablation study.
We observe that all studied components contribute to the effectiveness of our model to some degree.
Among all the components, removing pre-trained embeddings and removing feature smoothing steps cause larger performance drops compared to other components on the WeChat dataset.
In contrast, adding second-order features contributes a little to predict user behavior in WeChat \service, 
but contributes more to the Weibo dataset.
Meanwhile, our model performs well, even without hand-crafted user features.

\hide{
Table \ref{tb:like-prediction-var} and Table \ref{tb:click-prection-var} show the performance of our model variants on wow and click behavior prediction respectively. 
``-Features'' means removing customized user features (i.e. demographics and social role features) of input layers, and ``-GF'' means removing the step of feature smoothing via graph filter. 
From tables, we can infer that the learned user embeddings through modulated spectral graph domain provide better input node features which result in better prediction performance on F1 and AUC.
Moreover, we find that our model can perform well, even without input hand-crafted user features.
}
\subsection{Parameter Analysis}

\hide{
\begin{figure}[t]
	\centering
	\hspace*{0cm}
	\hspace{-0.05in}
	\mbox{
		\subfigure[Wow Performance]{
			\includegraphics[width=4.2cm]{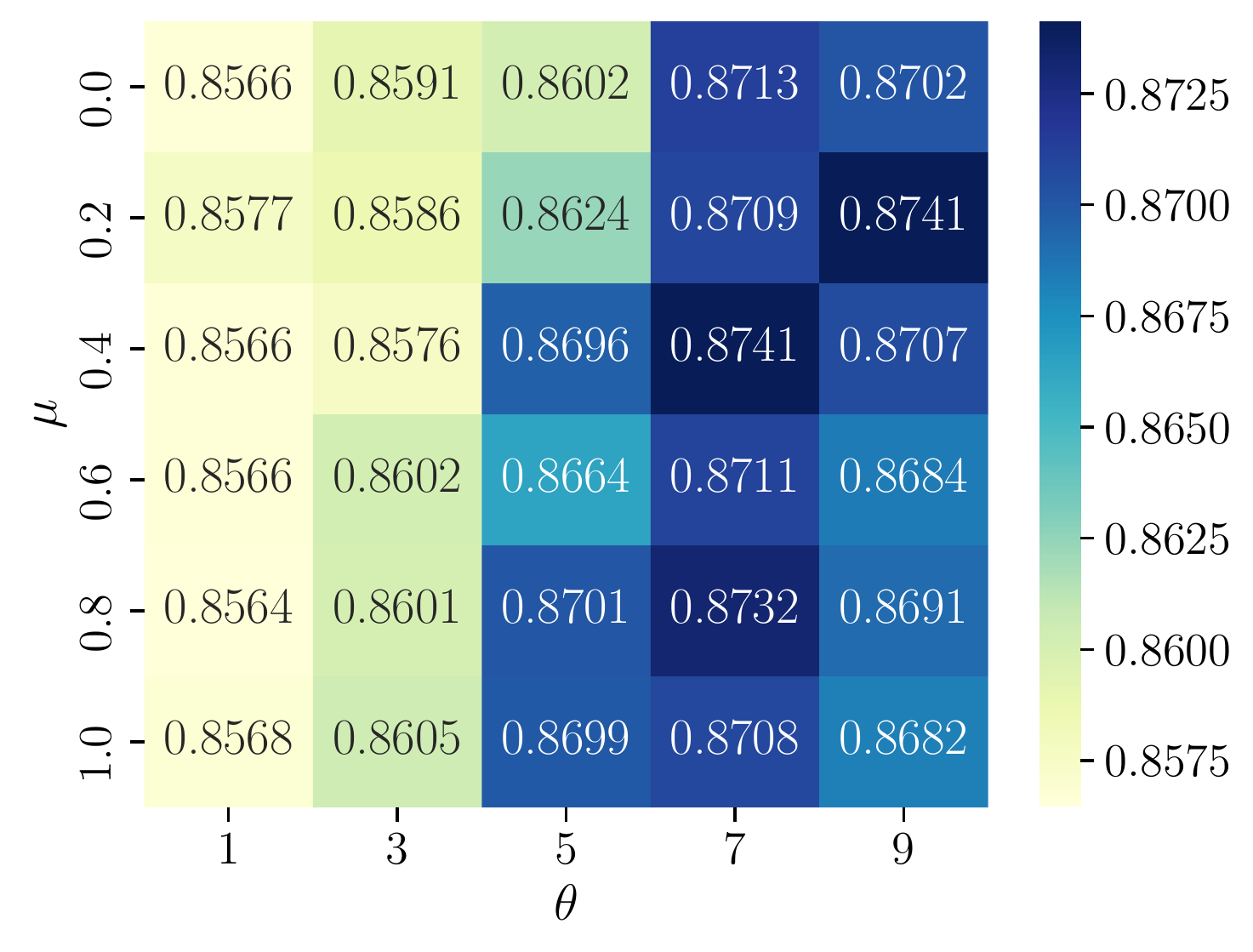}	
			\label{figsub:mu-theta-wow}
		}
		\subfigure[Click Performance]{
			\includegraphics[width=4.2cm]{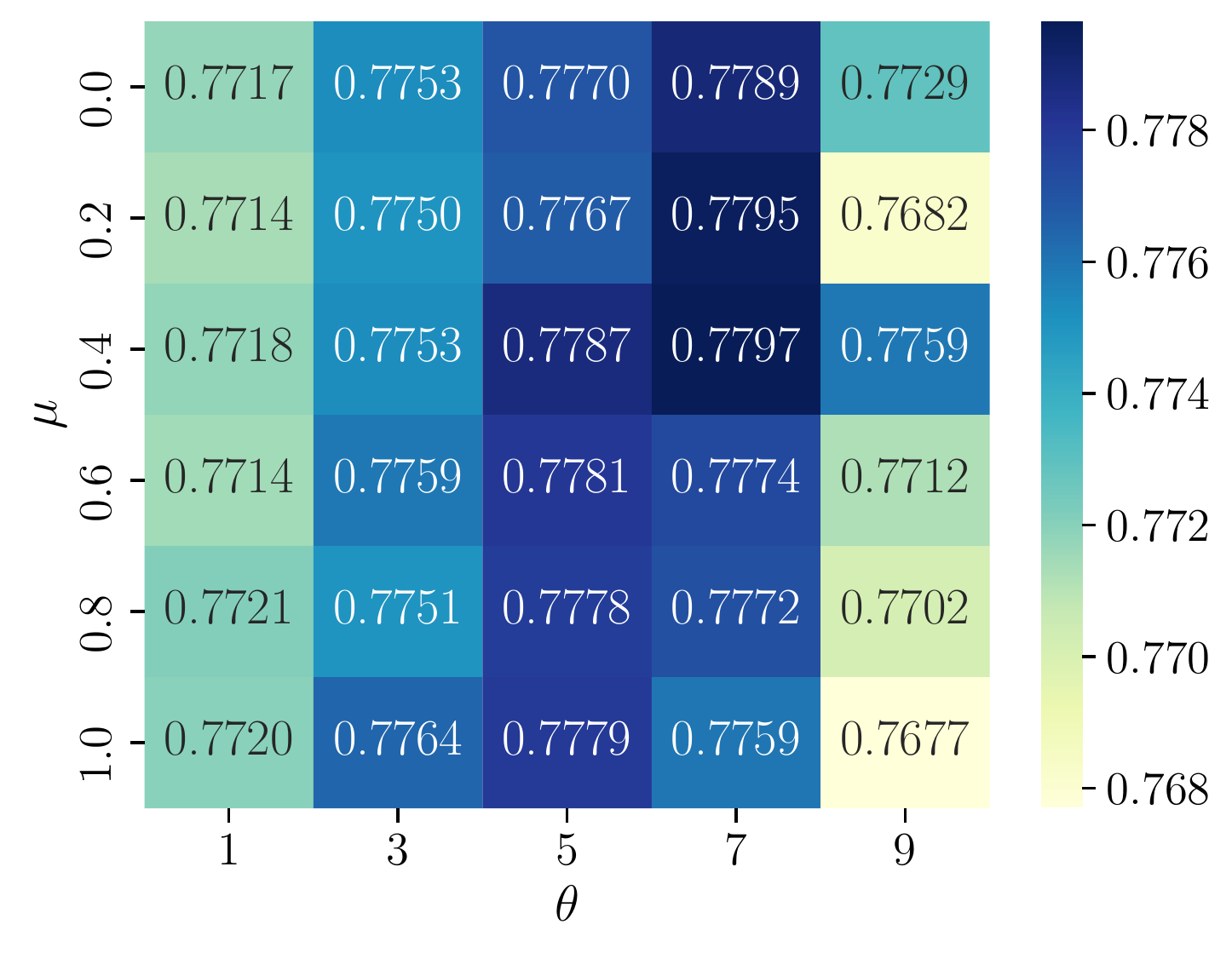}	
			\label{figsub:mu-theta-click}
		}
	}
	\vspace{-0.13in}
	\caption{Wow and click performance (AUC) on test dataset w.r.t. $\mu$ and $\theta$ in the graph filter
	}
\end{figure}
}

\vpara{The Number of Coarsening Steps in Hierarchical Graph Representation Learning.}
In the hierarchical graph representation step, each coarsening step transforms the ego network to smaller graphs iteratively by clustering similar ``nodes'' together. We study whether the number of coarsening steps influences the prediction performance. Figure \ref{fig:num-pooling-effect} shows the prediction performance w.r.t. the number of coarsening steps.
In the experiment, we set in each iteration, the number of ``nodes'' becomes half of that of the last iteration.
We see that hierarchical graph representation clearly better than ``flat'' representation (0 pooling layer) on WeChat ``Wow'' and Weibo dataset.
The prediction performance changes little in terms of the number of coarsening steps for click behavior prediction.
When there are 2 coarsening steps, test AUC and F1 for click prediction are the highest.
1 pooling layer is the best for the Weibo dataset and WeChat ``wow''.
Perhaps although we set the coarsening ratio to 50\%, GNN can automatically learn the node proximity and the appropriate number of meaningful clusters~\cite{ying2018hierarchical}.

\vpara{Parameters in the graph filter.} We analyze how the parameters  $\theta$ in the graph filter $g(\lambda) = e^{-\frac{1}{2} [(\lambda - \mu)^2 - 1] \theta}$ can affect the prediction performance. Here $\mu$ is trainable and we set its initial value to 0.4. Figure \ref{fig:theta-performance} shows the performance variations (AUC and F1) w.r.t. $\theta$ on test data.
Parameter $\theta$ in $g$ affects the peak value of modulated eigenvalue $\lambda$.
We observe that when $\theta = 7$ or $\theta = 9$, 
AUC is higher for ``wow'' and click prediction than other tested configurations.
However, for the Weibo dataset, when $\theta$ varies from 1 to 9, test performance increases first, then decreases. 

\hide{
\begin{figure}[t]
	\centering
	\includegraphics[width=8cm]{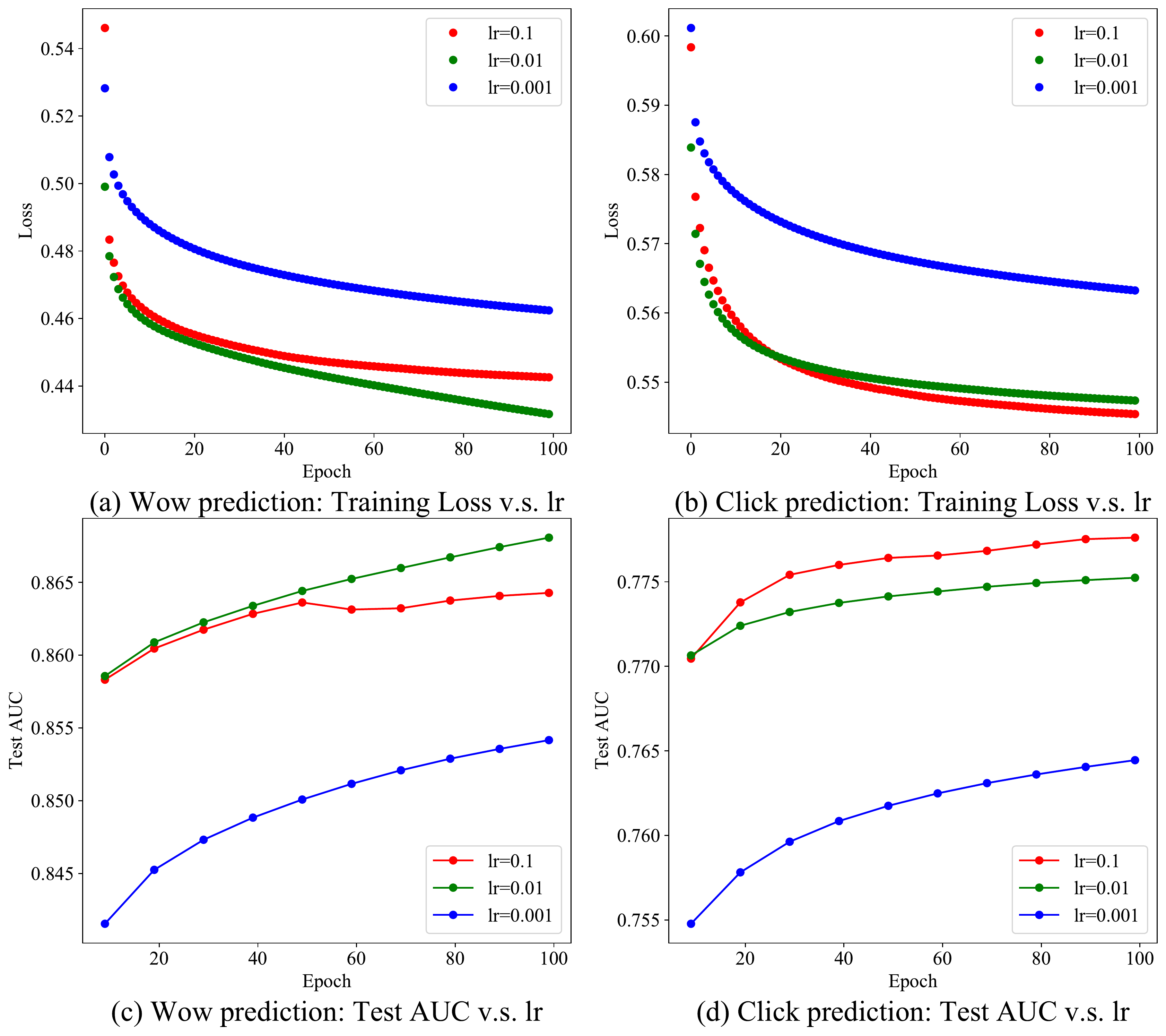}
	\vspace{-0.5cm}
	\caption{The training loss and test AUC v.s. different learning rates (lr).}
	\label{fig:lr-loss-auc}
	\vspace{0.5cm}
\end{figure}
}

\hide{
\vpara{Learning Rate.}
We analyze how to set an appropriate learning rate to benefit the convergence of the training process and prediction performance.
Figure \ref{fig:lr-loss-auc} shows the training loss and test AUC of ``wow'' and click prediction with the learning rate (lr) varied in the range of \{0.1, 0.01, 0.001\}.
From Figure \ref{fig:lr-loss-auc} (a) and (c), we observe that setting $lr = 0.01$ is suitable for ``wow'' prediction. 
As shown in Figure \ref{fig:lr-loss-auc} (b) and (d), click prediction is a little bit different, with the
best configuration $lr = 0.1$.
A possible reason is that large learning enables the learner to jump out of the local minimum. 
Comparing the training loss between ``wow'' and click prediction, the converging loss of click prediction is higher than that of ``wow'' prediction, which demonstrates the click behavior is harder to predict for our method.
}

\subsection{Visualization of Hierarchical Graph Representation Learning}

\begin{figure}[t]
	\centering
	\hspace*{0cm}
	\hspace{-0.05in}
	\mbox{
		\subfigure[Positive ``wow'' sample]{
			\includegraphics[width=4cm]{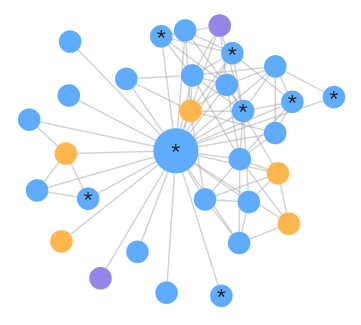}	
			\label{figsub:cluster-assign-wow-pos}
		}
		\hspace{0.1in}
		\subfigure[Negative ``wow'' sample]{
			\includegraphics[width=4cm]{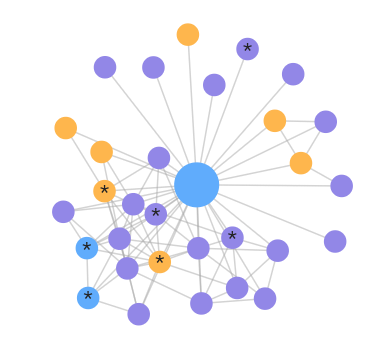}	
			\label{figsub:cluster-assign-wow-neg}
		}
	}
	\vspace{-0.13in}
	\caption{Visualization of cluster assignment of hierarchical graph representation learning for ``wow'' behavior prediction. (a) Positive sample: the ego user ``wow''ed the article. (b) Negative sample: the ego user didn't ``wow'' the article.
	Here different colors represent different cluster assignments.
		The largest node in each ego network is the ego user.
		Nodes with ``*'' inside mean that these users ``wow''ed the article.
	}
\end{figure}

We visualize the hierarchical graph representation learning process to understand how it assigns nodes into different clusters and performs predictions.
Figure \ref{figsub:cluster-assign-wow-pos} and \ref{figsub:cluster-assign-wow-neg} show two case studies for ``wow'' prediction,
where  Figure \ref{figsub:cluster-assign-wow-pos} is a positive ``wow'' instance and  Figure \ref{figsub:cluster-assign-wow-neg} is a negative ``wow'' instance.
The two subfigures both visualize the first coarsening step.
We observe that, although the coarsening ratio is set to 50\%, which means the nodes in one ego network can be assigned to at most 16 clusters, 
the two examples both assign nodes into only three clusters.
It shows that the learning algorithm can automatically learn the appropriate number of clusters. Moreover, for the positive ``wow''  instance in Figure \ref{figsub:cluster-assign-wow-pos}, all active friends (who ``wow''ed the article) are assigned to blue clusters, which may correspond to the conformity phenomenon.
As for the negative ``wow'' instance in Figure \ref{figsub:cluster-assign-wow-neg}, active friends are assigned to different clusters (nodes with different colors all have nodes with ``*'' within).
The inactivity of the ego user perhaps can be explained by the negative correlation between ego users' ``wow'' probability and structural diversity (referring to Sec. \ref{sec:ego-net-property}).
\section{Related Work}

\subsection{Social Influence Analysis}
Social influence has been studied and modeled widely from different viewpoints. At the macro level, the problem of influence maximization in social networks has been studied in ~\cite{David2003maximize,doming2001netvalue}. Xin et al.~\cite{Xin2012influence} study the indirect influence on Twitter. Micro influence, like pairwise influence, has been studied in~\cite{Goyal2010Influence,Saito2008Influence,zhang2013social}. Liu et al.~\cite{Liu2012influence} study the micro mechanism of influence diffusion in heterogeneous social networks and propose a probabilistic generative model. Tang et al.~\cite{tang2009social} propose Topical Affinity Propagation (TAP) to model influence on different topics in the academic network~\cite{tang2008arnetminer} and the heterogeneous network. 
More recently, deep learning models have been proposed to model social influence. Qiu et al.~\cite{qiu2018deepinf} use Graph Attention Networks (GAT) to model social influence locality. Feng et al.~\cite{feng2018inf2vec} propose a skip-gram architecture to learn user embeddings that reflect social influence. 
In this work, we first analyze micro-level social influence (e.g., 
dyadic and triadic correlations) 
and local network structure, based on which we propose an effective method to model social influence.

\subsection{User Feedback in Recommender Systems}
Generally, user feedback in the recommender system falls into two categories: explicit feedback and implicit feedback. Implicit feedback includes click, mouse movement, etc., and explicit feedback includes retweets, ratings, and so on. Jawaheer et al.~\cite{jawaheer2014modeling} propose a classification framework for explicit and implicit feedback based on several properties, including Cognitive Effort, User Model, Scale of Measurement, and Domain Relevance. 
They also compare different user feedback types in detail in an online music recommendation service~\cite{jawaheer2010comparison}. 
Many recommender systems try to combine different kinds of user feedback to improve recommendation performance. Liu et al.~\cite{liu2010unifying} unify explicit and implicit feedback in a matrix-factorization framework. However, Tang et al.~\cite{tang2016empirical} find that it is better to train different models separately for each feedback type and then combine them. 
In this work, we analyze both ``wow'' and click user feedback, and discover some interesting differences between them. Then we train separate models for each feedback type.

\subsection{Network Representation Learning}
Recently, network representation learning at the \textit{node level} and \textit{graph level} has become a research hotspot.
Generally, node-level representation learning approaches can be broadly categorized as (1) factorization-based approaches 
such as GraRep~\cite{cao2015grarep}, NetMF\cite{qiu2018network},
(2) shallow embedding approaches
such as DeepWalk~\cite{perozzi2014deepwalk}, LINE~\cite{tang2015line}, 
HARP~\cite{chen2018harp}, 
and (3) neural network approaches~\cite{cao2016deep, li2018adaptive}. Recently, graph convolutional network (GCN)~\cite{kipf2017semi} and its multiple variants, 
such as GAT~\cite{velivckovic2018graph}, GIN~\cite{xu2018powerful}, have become the dominant approaches for network representation learning, thanks to the use of graph convolution that effectively fuses graph topology and node features. 
Furthermore, there are also some works using graph convolution architectures for graph-level representation learning, such as~\cite{ying2018hierarchical, lee2019self, sun2019infograph}. In order to generate graph representations, most works employ graph convolution encoders to generate node embeddings first, and then use some pooling or READOUT functions, such as hierarchical pooling~\cite{ying2018hierarchical} 
and sum operations~\cite{sun2019infograph}.
Among hierarchical pooling methods, some~\cite{lee2019self,ranjan2020asap} use self-attention to select important nodes in the graph or cluster similar nodes together.
Yuan et al.~\cite{yuan2020structpool} regard the node clustering problem as a structured prediction problem via conditional random fields.
In this work, based on users' ego networks, we first learn user/node embeddings by modulating the spectral domain of the ego networks. Then, a hierarchical graph representation method is utilized to generate graph-level embeddings. Our method is motivated by and consistent with our statistical analysis.
\section{Conclusion}
In this work, we use the WeChat Top Stories data to 
understand user preferences and ``wow'' diffusion.
Our study reveals several interesting phenomena:
1) Males' click probability is higher than females', while females' ``wow'' probability is higher than males'.
2) The active rate of young generations (users in their 20s) is the lowest.
3) Given the fixed number of friends who ``wow''ed an article, the larger \#CC (the number of connected components formed by active friends), the lower the ``wow'' probability of ego users is, but the higher the click probability is.

Based on these important discoveries, we also develop a unified
model \model to predict users’ behaviors. We evaluate
it on the real sizable social networks, and results show that the
proposed model can achieve significantly better performance over several state-of-the-arts.



%



\ifCLASSOPTIONcompsoc
  \section*{Acknowledgments}
\else
  \section*{Acknowledgment}
\fi

This work is supported by a research fund of Tsinghua-Tencent Joint Laboratory for Internet Innovation Technology, 
the
National Key R\&D Program of China (2018YFB1402600),
NSFC for Distinguished Young Scholar 
(61825602),
NSFC (61836013), 
NSFC (61672313),
%
and NSF under grants III-1763325, III-1909323, and SaTC-1930941. 

\ifCLASSOPTIONcaptionsoff
  \newpage
\fi



%

\bibliographystyle{abbrv}
\bibliography{ref}

\hide{

}

%

\begin{IEEEbiography}
	[{\includegraphics[width=1in,height=1.25in,clip,keepaspectratio]
		{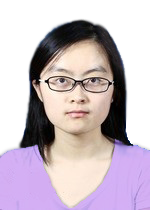}}]{Fanjin Zhang}
is a PhD candidate in the Department of Computer Science and Technology, Tsinghua University. She got her bachelor's degree from the Department of Computer Science and Technology, Nanjing University.
Her research interests include data mining and social networks.
\end{IEEEbiography}
\vspace{-0.5in}
\begin{IEEEbiography}[{\vspace{-5mm}\includegraphics[width=0.9in,height=1.25in,clip,keepaspectratio]{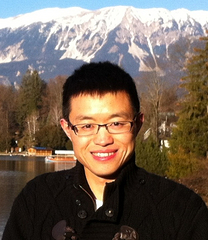}}]{Jie Tang}
is a Professor and the Associate Chair of the Department of Computer Science at Tsinghua University. He is a Fellow of the IEEE. His interests include artificial intelligence, data mining, social networks, and machine learning. He was honored with the SIGKDD Test-of-Time Award, the UK Royal Society-Newton Advanced Fellowship Award, NSFC for Distinguished Young Scholar, and KDD’18 Service Award.
\end{IEEEbiography}
\vspace{-0.5in}
\begin{IEEEbiography}[{\vspace{-3mm} \includegraphics[width=1in,height=1.25in,clip,keepaspectratio]{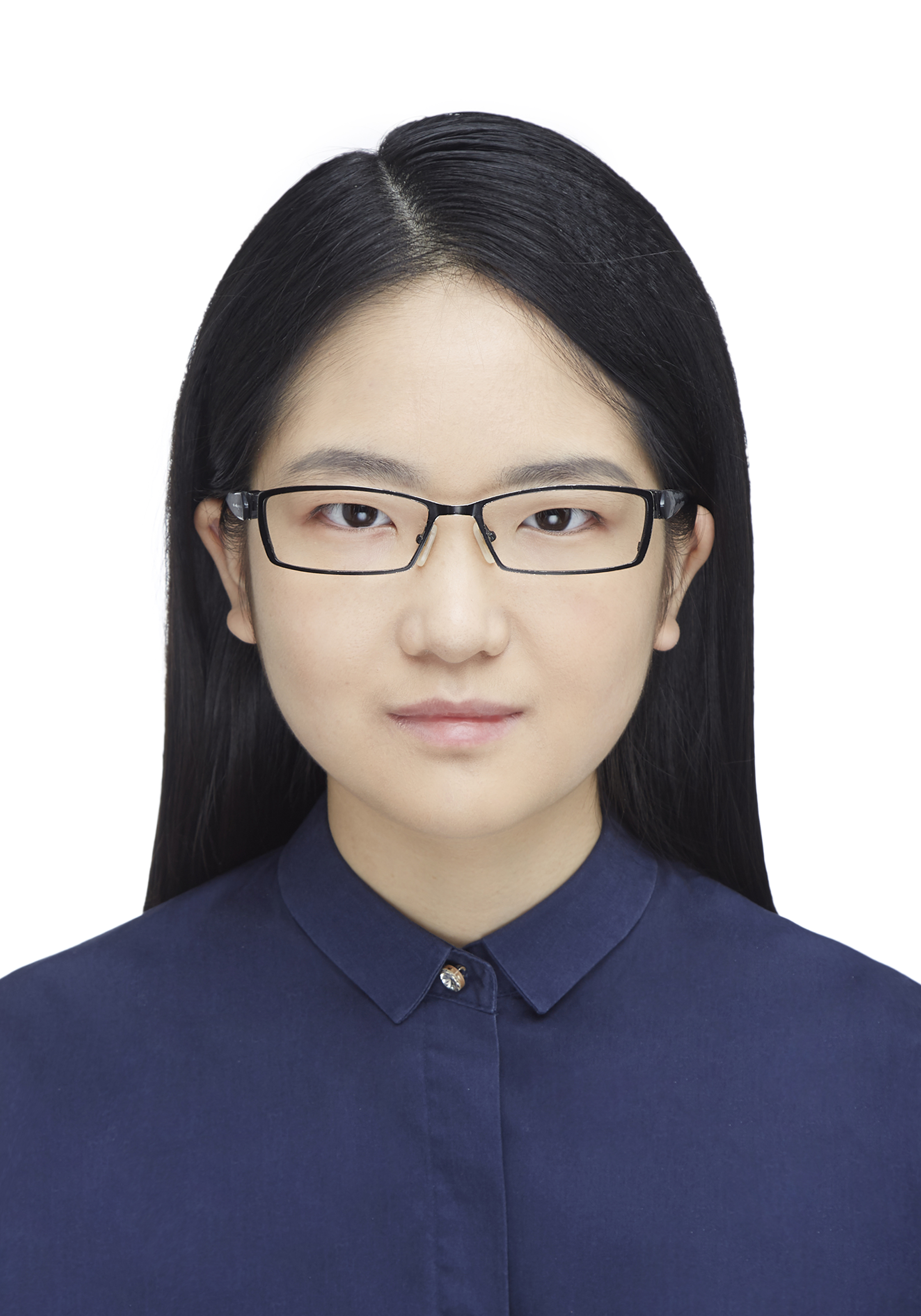}}]{Xueyi Liu}
is an undergraduate with the department of Computer Science and Technology, Tsinghua University. Her main research interests include graph representation learning, graph neural networks and reasoning.
\end{IEEEbiography}
\vspace{-0.5in}
\begin{IEEEbiography}[{\vspace{-5mm} \includegraphics[width=1in,height=1.25in,clip,keepaspectratio]{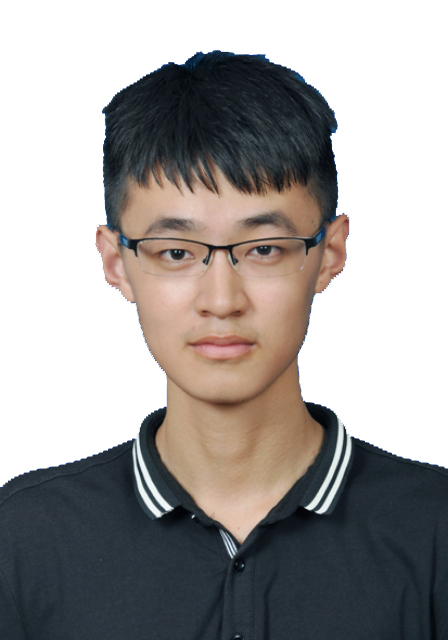}}]{Zhenyu Hou}
is an undergraduate with the department of Computer Science and Technology, Tsinghua University. His main research interests include graph representation learning and reasoning.
\end{IEEEbiography}
\vspace{-0.5in}
\begin{IEEEbiography}[{\vspace{-8.5mm} \includegraphics[width=0.9in,height=1.25in,clip,keepaspectratio]{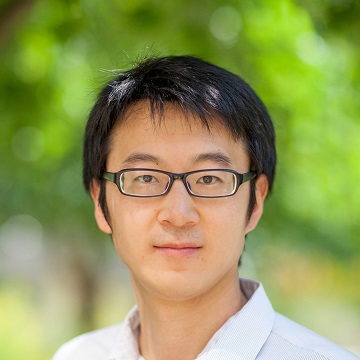}}]{Yuxiao Dong} 
received his Ph.D. from University of Notre Dame in 2017.
He is a Senior Researcher at Microsoft Research
Redmond. His research focuses on graph representation learning, knowledge graphs, social \& information networks, and data mining, with an emphasis on developing machine learning models to address problems in Web-scale systems.
\end{IEEEbiography}
\vspace{-0.5in}
\begin{IEEEbiography}[{ \vspace{-7mm} \includegraphics[width=0.9in,height=1.25in,clip,keepaspectratio]{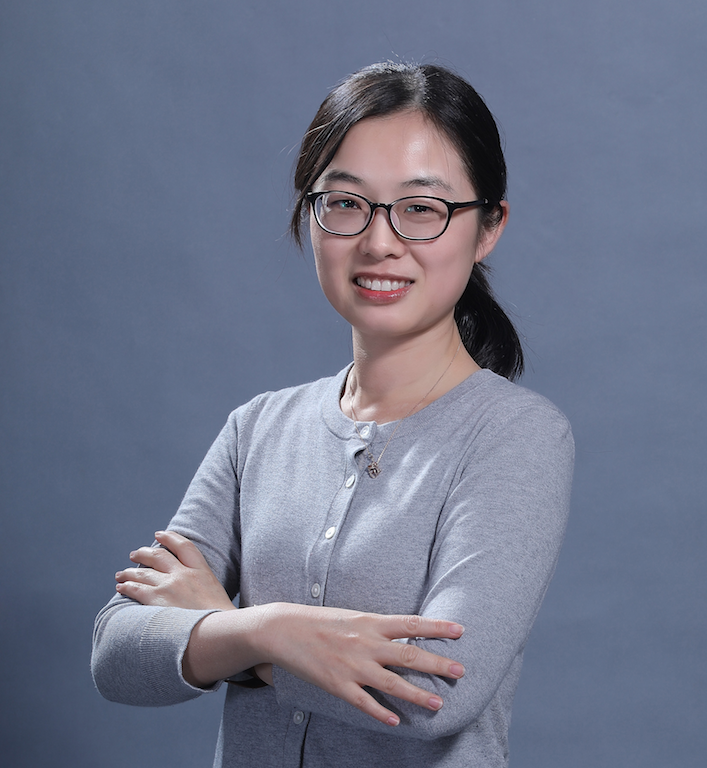}}]{Jing Zhang}
	received the master and PhD
	degrees from the Department of Computer
	Science and Technology, Tsinghua University. She is an associate professor with the Information School, Renmin University of China.
	Her research interests include social network
	mining and deep learning.
\end{IEEEbiography}
\vspace{-0.5in}
\begin{IEEEbiography}[{\vspace{-4mm}\includegraphics[width=0.9in,height=1.25in,clip,keepaspectratio]{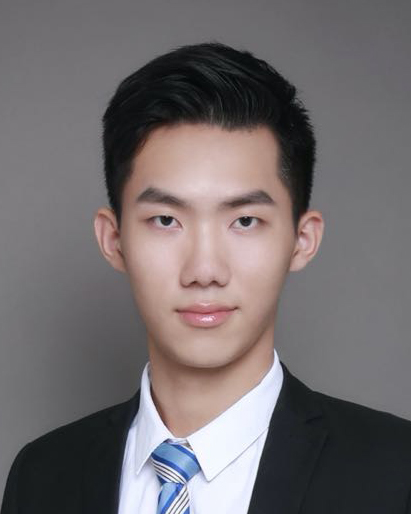}}]{Xiao Liu}
	is a senior undergraduate student with the Department of Computer Science and Technology, Tsinghua University. His main research interests include data mining, machine learning, and knowledge graph. He has published papers on KDD.
\end{IEEEbiography}
\begin{IEEEbiography}[{\vspace{-1mm}\includegraphics[width=0.9in,height=1.25in,clip,keepaspectratio]{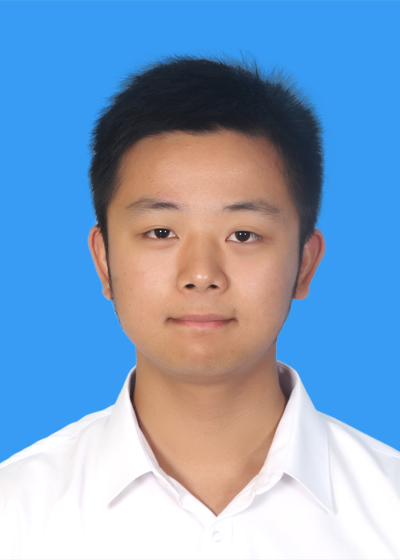}}]{Ruobing Xie}
is a researcher of WeChat, Tencent. He got his BEng degree in 2014 and his master degree in 2017 from the Department of Computer Science and Technology, Tsinghua University. His research interests are natural language processing and recommender system. He has published over 15 papers in international journals and conferences including IJCAI, AAAI, ACL and EMNLP.
\end{IEEEbiography}
\vspace{-2.6in}
\begin{IEEEbiography}	[\vspace{-2mm} {\includegraphics[width=0.9in,height=1.25in,clip,keepaspectratio]{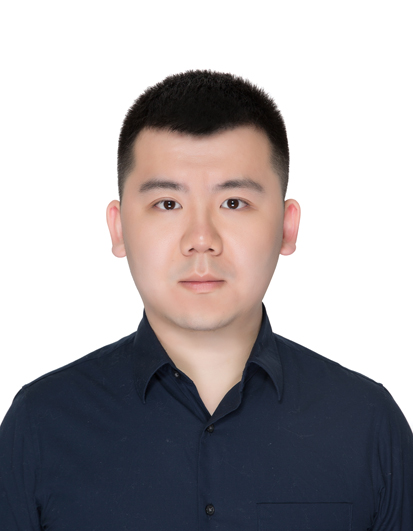}}]
	{Kai Zhuang}
	received his master degree from the
	Department of Computer Science and Technology, Sichuan University, China, in 2013. He
	is a researcher in WeChat Search Application Department, Tencent Inc., China.
\end{IEEEbiography}
\vspace{-2.6in}
\begin{IEEEbiography}[{\vspace{-12mm}\includegraphics[width=0.9in,height=1.25in,clip,keepaspectratio]{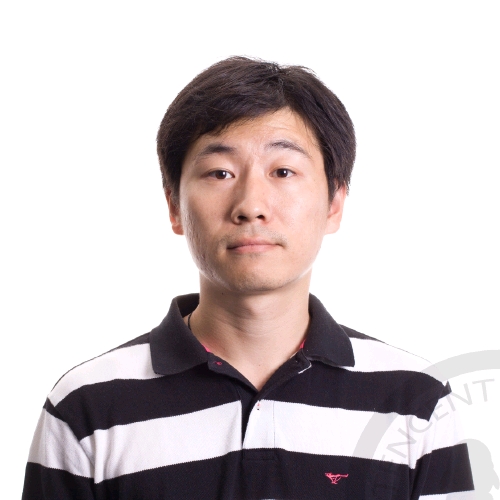}}]{Xu Zhang}
received his master degree in the College of Computer Science and Engineering, Northeastern University in 2008. He is the team leader of WeChat User Profile, Tencent. His research interests include machine learning and its applications, such as social networks, user interest models and user behavior research.
\end{IEEEbiography}
\vspace{-2.6in}
\begin{IEEEbiography}[{\vspace{-12mm}\includegraphics[width=0.9in,height=1.25in,clip,keepaspectratio]{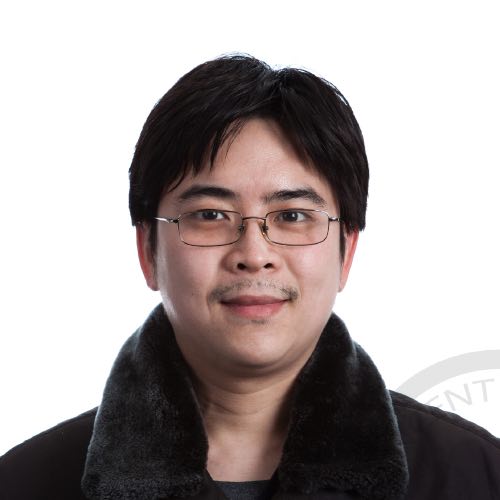}}]{Leyu Lin}
received his master degree from the Institute of Computing Technology, Chinese Academy of Sciences, in 2008. He is currently the deputy director of WeChat Search Team, Tencent. His research interests include machine learning and its applications, such as search system, recommendation systems and computational advertising.
\end{IEEEbiography}
\vspace{-2.6in}
\begin{IEEEbiography}[{\vspace{-4mm}\includegraphics[width=0.9in,height=1.25in,clip,keepaspectratio]{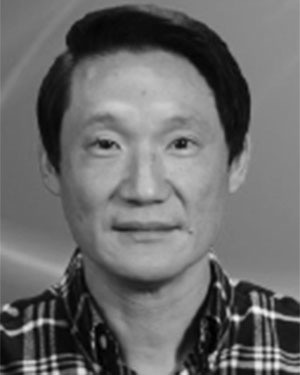}}]{Philip S. Yu}
received the PhD degree in electrical
engineering from Stanford University. He is a distinguished
professor in computer science with the
University of Illinois, Chicago, and is the Wexler
chair in Information Technology. His research interests
include big data, data mining, data stream,
database, and privacy. He was the editor-in-chief of
the IEEE Transactions on Knowledge and Data
Engineering and the ACM Transactions on Knowledge
Discovery from Data. He received the ACM
SIGKDD 2016 Innovation Award, a Research Contributions
Award from the IEEE International Conference on Data Mining
(2003), and a Technical Achievement Award from the IEEE Computer Society
(2013). He is a fellow of the IEEE and ACM.
\end{IEEEbiography}




\end{document}